\pdfoutput=1
\documentclass[11pt]{amsart}
\usepackage{fullpage}
\usepackage{amssymb}
\usepackage[ruled]{algorithm}
\usepackage{algorithmic}
\usepackage{graphics}
\usepackage{latexsym}
\usepackage{amsfonts}
\usepackage{boxedminipage}
\usepackage{xspace}
\usepackage{longtable}
\usepackage{tikz}
\usepackage{color}
\usepackage{paralist}
\usepackage{fancyvrb}

\newcommand{\mts}{Max 2-Sat\xspace}
\newcommand{\mtc}{Max (2,2)-CSP\xspace}
\newcommand{\mc}{Max 2-CSP\xspace}

\newcommand{\myNot}[1]{\overline{#1}\xspace}

\newcommand{\wlg}{without loss of generality\xspace}

\newtheorem{theorem}{Theorem}
\newtheorem{lemma}[theorem]{Lemma}

\newtheorem{remark}[theorem]{Remark}

\newcommand{\bool}{\{0,1\}}
\newcommand{\ZZ}{\mathbb{Z}}
\newcommand{\Ostar}[1]{{O^\star}\left( {#1} \right)}

\newcommand{\set}[1]{\{ #1 \}}
\def\twomu(#1){2^{\mu(#1)}}
\newcommand{\twopow}[1]{2^{#1}}

\newcommand{\OR}{\lor}
\newcommand{\AND}{\land}
\newcommand{\XOR}{\oplus}

\newcommand{\bias}{\operatorname{bias}}
\newcommand{\myitem}[1]{\subsection{#1}}
\newcommand{\myref}[1]{(reduction~\ref{#1})}
\def\hisitem[#1]{\subsection{#1}}

\newcommand{\poly}{\operatorname{poly}}
\def\polyk(#1){ {{#1}} ^k }
\newcommand{\proofsymbol}{$\Box$}
\newcommand{\casesymbol}{$\boxdot$}
\newcommand{\caseend}{\hspace*{1em} \hfill \casesymbol}
\DeclareMathOperator*{\maxdeg}{maxdeg}
\newcommand{\ceil}[1]{\left\lceil {#1} \right\rceil}
\newcommand{\floor}[1]{\left\lfloor {#1} \right\rfloor}

\newcommand{\murest}{
 \sum_{d=4}^6 \chi(\maxdeg(G) \geq d) C_d 
  + \sum_{d=4}^6 \chi(G \text{ is $d$-regular}) R_d
}

\newcommand{\constrain}[1]{#1}

\newcommand{\mua}{\nu}
\newcommand{\mub}{\delta}

\newcommand{\alga}{Algorithm~\ref{mainalg}\xspace}
\newcommand{\algb}{Procedure~\ref{simplifyalg}\xspace}

\newcommand{\reg}{r}
\newcommand{\nonreg}{{\overline{r}}}
\newcommand{\parens}[1]{\left( {#1} \right)}
\newcommand{\ssum}{\textstyle\sum}
\newcommand{\FF}{\mathcal{F}}

\title[A fast algorithm for Max 2-Sat]
{A universally fastest algorithm for Max 2-Sat, \\ Max 2-CSP,
and everything in between$^\dagger$}

\author[Serge Gaspers]{Serge Gaspers}

\address[Serge Gaspers]{
LIRMM -- University of Montpellier 2, CNRS\\
34392 Montpellier\\
France}
\email{gaspers@lirmm.fr}

\author[Gregory B. Sorkin]{Gregory B. Sorkin}

\thanks{$^\dagger$ 
The research was done largely during Serge Gaspers' visit 
to IBM Research in July--September 2007,
and Gregory Sorkin's visit to the University of Bergen 
in April 2008, both with support from the Norwegian Research Council.\\
A preliminary version of this paper appeared in the proceedings of SODA 2009~\cite{GaspersSorkin09}.
}

\address[Gregory B. Sorkin]{
Department of Mathematical Sciences \\
IBM T.J.\ Watson Research Center \\
Yorktown Heights NY 10598, USA}
\email{sorkin@watson.ibm.com}

\begin{document}

\bibliographystyle{amsalpha}

\date{June 2009}

\begin{abstract}
In this paper we introduce ``hybrid'' \mc formulas 
consisting of ``simple clauses'', 
namely conjunctions and disjunctions of pairs of variables, 
and general 2-variable clauses,
which can be any integer-valued functions of pairs of boolean variables.
This allows an algorithm to use both efficient reductions specific to AND and OR clauses,
and other powerful reductions that require the general CSP setting.
We use new reductions introduced here,
and recent reductions such as ``clause-learning'' and ``2-reductions''
generalized to our setting's mixture of simple and general clauses.

Parametrizing an instance by the fraction $p$ of non-simple clauses,
we give an exact (exponential-time) algorithm that is 
the fastest known polynomial-space algorithm for $p=0$
(which includes the well-studied \mts problem 
but also instances with arbitrary mixtures of AND and OR clauses);
the only efficient algorithm for mixtures of AND, OR,
and general integer-valued clauses;
and tied for fastest for general \mc ($p=1$).
Since a pure 2-Sat input instance may be transformed to 
a general CSP instance in the course of being solved,
the algorithm's efficiency and generality go hand in hand.

Our algorithm analysis and optimization are a variation on the 
familiar measure-and-conquer approach,
resulting in an optimizing mathematical program that is
convex not merely quasi-convex,
and thus can be solved efficiently and with a certificate of optimality.
We produce a {family} of running-time upper-bound formulas,
each optimized for instances with a particular value of~$p$
but valid for all instances. 
\end{abstract}

\maketitle

\tableofcontents

\section{Introduction}
\subsection{Treatment of ``hybrid'' Sat--CSP formulas}
We show a polynomial-space algorithm that solves general instances of integer-valued \mc
(formally defined in Section~\ref{sec:pre}),
but that takes advantage of ``simple'' clauses,
namely unit-weighted conjunctions and disjunctions,
to reduce the running time.
In a sense made precise near Remark~\ref{booleanfunctions}, 
exclusive-or is the only boolean function we cannot treat efficiently.

Let us give a simple example. In the \mc instance
\begin{equation}\label{max2cspExample}
 (x_1 \OR x_2) + (x_2 \OR \myNot{x_4}) + (x_2 \AND x_3) + 3 \cdot (x_1 \OR x_3) + (2 \cdot (\myNot{x_2}) - 5 \cdot x_4 + (x_2 \XOR x_4)),
\end{equation}
the first two clauses are unit-weighted disjunctive clauses, the third clause is
a unit-weighted conjunction, the fourth clause is a disjunction with weight 3,
and the last clause is a general integer-valued CSP clause (any integer-valued 2-by-2 truth table).
Thus this example has 3 (the first three clauses) simple clauses and 2 non-simple clauses,
for a fraction of non-simple clauses of $p=2/5$.

\begin{table}
 \begin{tabular}{|llll|}
  \hline
  Running Time & Problem & Space & Reference\\\hline
  $\Ostar{\twopow{m/2.879}}$ & \mts & polynomial & Niedermeier and Rossmanith \cite{NiRo00}\\
  $\Ostar{\twopow{m/3.448}}$ & \mts & polynomial & implicit by Bansal and Raman \cite{BansalR99}\\
  $\Ostar{\twopow{m/4}}$ & \mts & polynomial & Hirsch \cite{Hirsch}\\
  $\Ostar{\twopow{m/5}}$ & \mts & polynomial & Gramm et al. \cite{Gramm03}\\
  $\Ostar{\twopow{m/5}}$ & \mc & polynomial & Scott and Sorkin \cite{random03} \\
  $\Ostar{\twopow{m/5.263}}$ & \mc & polynomial & Scott and Sorkin \cite{FasterIBM}\\
  $\Ostar{\twopow{m/5.217}}$ & \mts & polynomial & Kneis and Rossmanith \cite{kneis05}\\
  $\Ostar{\twopow{m/5.769}}$ & \mts & exponential & Kneis et al. \cite{Kneis4}\\
  $\Ostar{\twopow{m/5.5}}$ & \mts & polynomial & Kojenikov and Kulikov \cite{KK}\\
  $\Ostar{\twopow{m/5.769}}$ & \mc & exponential & Scott and Sorkin \cite{faster}\\
  $\Ostar{\twopow{m/5.88}}$ & \mts & polynomial & Kulikov and Kutzkov \cite{clauseLearning}\\
  $\Ostar{\twopow{m/6.215}}$ & \mts & polynomial & Raible and Fernau \cite{Raible}\\\hline
 \end{tabular}
 \caption{\label{tab:papers}A historical overview of algorithms for \mts and \mc}
\end{table}

Both \mts and \mc have been extensively studied from the algorithmic point of view.
For variable-exponential running times, the only two known algorithms faster than $\twopow{n}$ for \mc (and \mts)
are those by Williams~\cite{Williams05} and Koivisto~\cite{Koivisto06csp}, both with running time $\Ostar{\twopow{n/1.262}}$.
They employ beautiful ideas, but have exponential space complexity.

For clause-exponential running times, there has been a long series of improved algorithms; see Table~\ref{tab:papers}.
To solve \mts, all early algorithms treated pure 2-Sat formulas. By using more powerful reductions closed over \mc but not \mts,
the \mc generalization of Scott and Sorkin \cite{FasterIBM} led to a faster algorithm. Then, several new \mts specific reductions
once again gave the edge to algorithms addressing \mts instances particularly.

In this paper we get the best
of both worlds by using reductions specific to \mts (actually, we allow disjunctive and conjunctive clauses),
but also using CSP reductions. While it is likely that \mts algorithms will become still faster, we believe that further improvements will continue to use this method of combination.

\subsection{Results}
Let $p$ be the fraction of non-simple clauses in the initial instance, no matter how this fraction changes during the execution of the algorithm.
In Example \eqref{max2cspExample}, $p=2/5$.
The algorithm we present here is the fastest known polynomial-space algorithm for $p=0$
(including \mts but also instances with arbitrary mixtures
of AND and OR clauses);
fastest for $0 < p < 0.29$
(where indeed no other algorithm is known, short of solving the
instance as a case of general \mc);
and tied for fastest for $0.29 \leq p \leq 1$,
notably for \mc itself.
For the well-known classes \mts and \mc, our algorithm
has polynomial space complexity and
running time $\Ostar{\twopow{m/6.321}}$ and $\Ostar{\twopow{m/5.263}}$, respectively.

For ``cubic'' instances, 
where each variable appears in at most three 2-variable clauses,
our analysis gives running-time bounds 
that match and generalize the best known when $p=0$ (including \mts);
improve on the best known when $0<p<1/2$;
and match the best known for $1/2 \leq p \leq 1$ (including \mc).

We derive running-time bounds that are optimized to the fraction $p$ 
of non-simple clauses; see Table~\ref{tab:runtimes}.
Every such bound is valid for every formula, 
but the bound derived for one value of $p$ 
may not be the best possible for a formula with a different value.

\subsection{Method of analysis, and hybrid Sat--CSP formulas}

Since a fair amount of machinery will have to be introduced before we
can fully explain 
our analysis, 
let us first give a simplified overview of the method,
including some new aspects of it in our application.
Our algorithm reduces an instance to one or more smaller instances,
which are solved recursively to yield a solution to the original instance.
We view a \mc instance as a constraint graph $G=(V,E \cup H)$ where
vertices represent variables, the set of ``light'' edges $E$ represents simple
clauses and the set of ``heavy'' edges $H$ respresents general clauses.
The reductions are usually local and change the 
constraint graph's structure, and a related \emph{measure}, in a predictable way. 

For example,
if $G$ has
two degree-4 vertices sharing two simple clauses, 
a ``parallel-edge'' reduction 
replaces the two simple clauses with one general clause,
changing the vertices' degrees from 4 to~3, 
giving a new constraint graph~$G'$.
With the measure $\mu$ including weights $w_e$ and $w_h$ 
for each simple and general clause
(mnemonically, the subscripts refer to ``edges'' and ``heavy'' edges),
and weights $w_3$ and $w_4$ for each vertex of degree 3 and~4,
this reduction changes an instance's measure by
$\mu(G')-\mu(G) = -2 w_e + w_h - 2 w_4 + 2 w_3$.
An inductive proof of a running-time bound
$\Ostar{\twopow{\mu(G)}}$
will follow \emph{if the measure change is non-positive}.
Thus, we constrain that 
\begin{align*}
 -2 w_e + w_h - 2 w_4 + 2 w_3 & \leq 0 .
\end{align*}

An algorithm requires a set of reductions covering all instances:
there must always be some applicable reduction.
Just as above, each reduction imposes a constraint on the weights.
One reduction's constraint can weaken those of other reductions,
by limiting the cases in which they are applied. 
For example, if we prioritize parallel-edge reduction,
given as an example above (generalized to all degrees),
we may assume that other reductions act on graphs without parallel edges.
More usefully, ``cut'' reductions will allow us to assume that a graph has no small vertex cuts.
Reductions like this 
producing a single instance,
or any number of isomorphic instances,
yield linear constraints (as in \cite{random03,FasterIBM,faster});
reductions producing distinct instances yield nonlinear, 
convex constraints.

If a set of weights giving a measure $\mu$ satisfies all the constraints, 
the analysis results in a proof of a running-time bound
$\Ostar{\twopow{\mu(G)}}$ for an input instance~$G$. 
To get the best possible running-time bound 
subject to the constraints,
we wish to minimize $\mu(G)$.
To avoid looking at the full degree spectrum of~$G$, 
we constrain each vertex weight $w_d$ 
to be non-positive, and then ignore these terms,
resulting in a (possibly pessimistic) running-time bound
$\Ostar{\twopow{|E| w_e + |H| w_h}}$. 

\begin{figure}[tbp]
 \begin{centering}
  \includegraphics[width=0.7\textwidth,angle=-90]{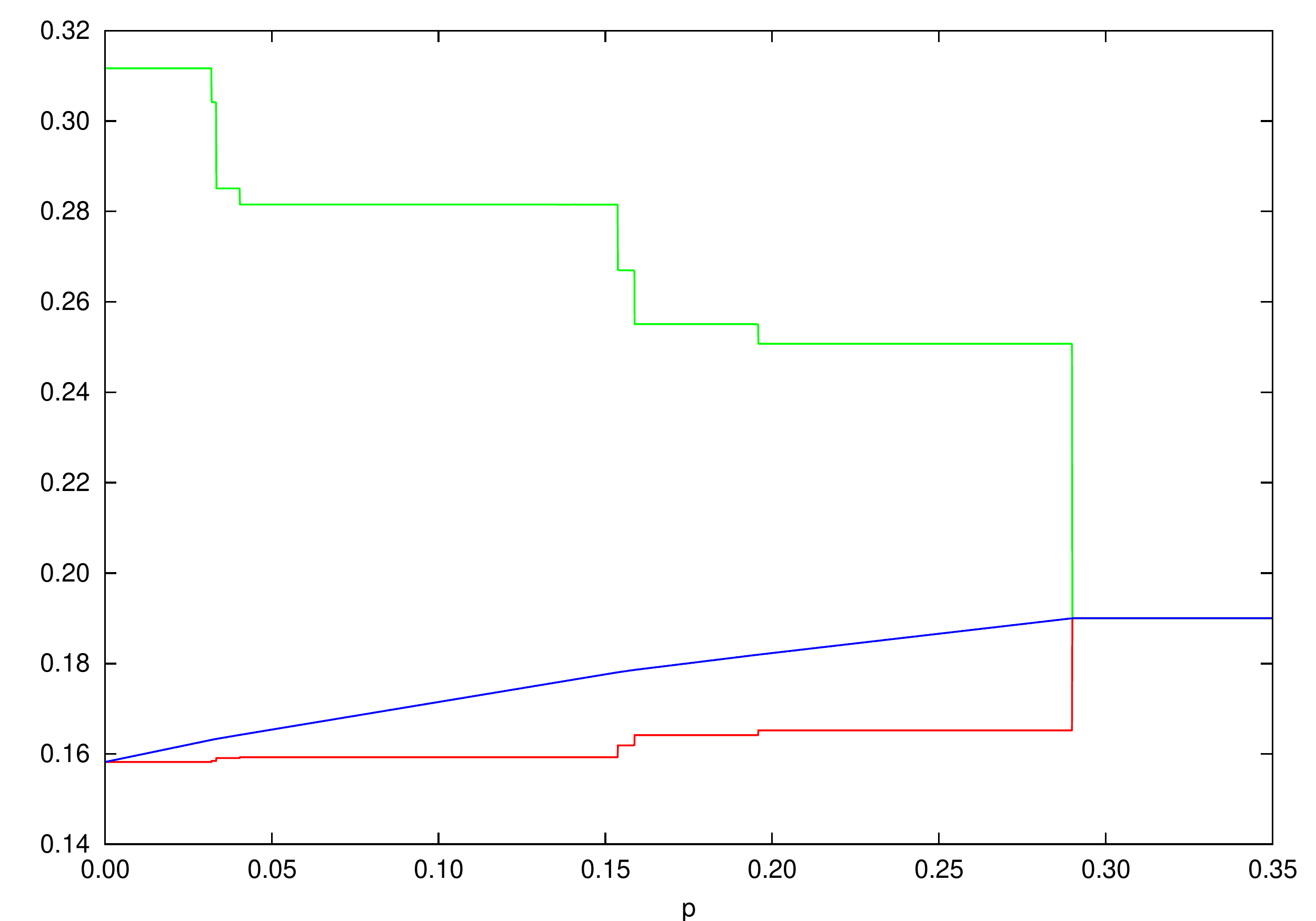}%
  \caption{Plot of $w_e$ (\textcolor{red}{red}), $w_h$ (\textcolor{green}{green}), and the running-time exponent $(1-p)w_e+p w_h$ (\textcolor{blue}{blue}) versus the fraction $p$ of non--simple 2-clauses. The three values are equal (and exactly 0.19) for $p>0.29$. Both $w_e$ and $w_h$ appear to be piecewise constant: the resolution of the graph is in $p$ increments of $0.0001$, and all the small changes are meaningful.\label{p6plot}}
\end{centering}
\end{figure}

If $G$ is a \mts instance, 
to minimize the running-time bound is simply to minimize $w_e$
subject to the constraints:
as there are no heavy edges in the input instance,
it makes no difference if $w_h$ is large.
This optimization will yield a small value of $w_e$ and a large $w_h$.
Symmetrically, if we are treating a general \mc instance,
where all edges are heavy, we need only minimize $w_h$.
This optimization will yield weights $w_e, w_h$ that are larger
than the \mts value of $w_e$ but smaller than its $w_h$.
For a hybrid instance with some edges of each type, 
minimizing $|E| w_e + |H| w_h$ is equivalent to 
minimizing $(1-p) w_e + p w_h$,
where $p=|H|/(|E|+|H|)$ is the fraction of non-simple clauses.
This will result in weights $w_e$ and $w_h$ each lying between 
the extremes given by the pure 2-Sat and pure CSP cases;
see Figure~\ref{p6plot}.

Thus, a new aspect of our approach is that it results in a family of
nonlinear programs (NLPs), not just one:
the NLPs differ in their objective functions, 
which are tuned to the fraction $p$ of non-simple clauses in an input instance.
The optimization done for a particular value of~$p$,
by construction, gives a running-time bound that is the best possible 
(within our methods)
for an input instance with this fraction of non-simple clauses,
but (because the constraints are the same in all the NLPs)
that is valid for all instances;
see the caption of Table~\ref{tab:runtimes}.

\subsection{Novel aspects of the analysis}
Our introduction of  the notion of hybrids between
\mts and \mc, discussed above,
is the main distinguishing feature of the present work.
It yields a more general algorithm, 
applicable to CSP instances not just Sat instances,
and gives better performance on \mts 
by allowing both efficient Sat-specific reductions
and powerful reductions that go outside that class.
This is surely not the final word on \mts algorithms, 
but we expect new algorithms to take advantage of this hybrid approach.

A secondary point is that CSP reductions such as 
combining parallel edges or reducing on small cuts 
mean that in other cases it can be assumed that a graph has 
no parallel edges or small cuts.
This potentially decreases the running-time bound (by weakening the corresponding constraints), and simplifies the case analysis,
counter-balancing the complications of considering two types of edges.

Our analysis uses a now-common method, but with some novel aspects.
Specifically, we analyze a reduction-based algorithm with 
a potential-function method akin to the 
measures used by~\cite{Kullmann2,Kullmann1},
the quasi-convex analysis of~\cite{Eppstein}, 
the ``measure and conquer'' approach of~\cite{conquer},
the (dual to the) linear programming approach of~\cite{faster},
and much older potential-function analyses in mathematics and physics.
The goal is to solve a NLP
giving a set of weights which minimizes a running-time bound, 
while respecting constraints imposed by the reductions.
The hybrid view marks one change to this approach, 
since, as already discussed, it means that the
objective function depends on the fraction of non-simple clauses,
so there is a continuum of NLPs, not just one.

Our nonlinear programs are convex
(those of \cite{Eppstein} are only quasi-convex),
allowing them to be solved quickly 
and with certificates of optimality.

Also, it is common to make some assumptions about the weights, 
but we try to avoid this,
instead only limiting the weights 
by the constraints necessitated by each reduction.
This avoids unnecessary assumptions compromising 
optimality of the result, 
which is especially important in the hybrid realm
where an assumption might be justified for Sat but not for CSP,
or vice-versa.
It also makes the analysis more transparent. 

As is often the case with exact algorithms, 
regularity of an instance is important, 
and in our analysis we treat this with explicit weights 
penalizing regularity
(motivated by a similar accounting for the number of 2-edges in 
a hypergraph in \cite{Wahlstrom2004},
and the ``forced moves'' in \cite{faster}).
This introduces some extra bookkeeping but results in a 
more structured, more verifiable analysis.

We introduce several new reductions, including 
a 2-reduction combining ideas from \cite{KK} (for the Sat case)
and \cite{faster} (the CSP case),
a ``super 2-reduction'',
and a generalization of the ``clause-learning'' from~\cite{clauseLearning}.

\section{Definitions}
\label{sec:pre}

We use the value 1 to indicate Boolean ``true'', and 0 ``false''.
The canonical problem Max Sat is,
given a boolean formula in conjunctive normal form (CNF), 
to find a boolean assignment to the variables of this formula
satisfying a maximum number of clauses.
\mts is Max Sat restricted to instances 
in which each clause contains at most 2 literals.

We will consider a class more general than \mts,
namely integer-valued \mtc; we will generally abbreviate this to \mc.
An instance $(G, S)$ of \mc is defined by a 
\emph{constraint graph} (or multigraph) $G = (V,E)$ and a set $S$ of
\emph{score} functions. 
There is a \emph{dyadic} score function 
$s_e \colon \bool^2 \to \mathbb{\ZZ}$ 
for each edge $e \in E$,
a monadic score function $s_v \colon \bool \to \mathbb{\ZZ}$ 
for each vertex $v \in V$, and (for bookkeeping convenience)
a single \emph{niladic} score ``function'' (really a constant)
$s_\emptyset \colon \bool^0 \to \mathbb{\ZZ}$.

A candidate solution is a function 
$\phi : V \to \bool$ assigning values to the vertices, 
and its score is
\[ s(\phi) :=  
  \sum_{uv \in E} s_{uv}(\phi(u), \phi(v))
  + \sum_{v \in V} s_v(\phi(v)) 
  + s_\emptyset .
\]
An optimal solution $\phi$ is one which maximizes $s(\phi)$.

The algorithm we present here solves any instance of \mc
with polynomial space usage, 
but runs faster for instances having a large proportion of 
``simple'' clauses, namely conjunctions and disjunctions.

A hybrid instance $F = (V, E, H, S)$ is defined by its 
variables or vertices~$V$, 
normal or \emph{light} edges $E$ representing conjunctive clauses
and disjunctive clauses, 
\emph{heavy} edges $H$ representing arbitrary (integer-valued) clauses, 
and a set $S$ of monadic functions and dyadic functions.
Its light-and-heavy-edged constraint graph is $G=(V,E,H)$,
though generally we will just think of the graph $(V,E \cup H)$;
no confusion should arise.
We will write $V(F)$ and $V(G)$ for the vertex set of an 
instance $F$ or equivalently that of its constraint graph~$G$.

In a graph $G$, we define the \emph{(open) neighborhood} of a vertex $u$
as $N(u) := \set{v: uv \in E \cup H} \setminus \set{u}$
(excluding $u$ will not matter once we simplify our
graphs and make them loopless),
and the {closed neighborhood} 
as $N[u] := N(u) \cup \set{u}$.
Generalizing, a \emph{set} of vertices, $U$, has
(open) {neighborhood} 
$N(U) = \left( \bigcup_{u \in U} N(u) \right) \setminus U$,
and (open) \emph{second neighborhood} 
$N^2(U) = N(N(U)) \setminus U$.
For a single vertex~$u$, define $N^2(u) := N^2(\set{u})$.
By definition, $U$, $N(U)$, and $N^2(U)$ are disjoint.

We define the degree $\deg(u)$ of a vertex $u$ to be the number of edges
incident on $u$ where loops are counted twice,
and the \emph{degree} (or \emph{maximum degree}) of a formula $F$
(or its constraint graph~$G$)
to be the maximum of its vertex degrees.
Without loss of generality we will assume that there is 
at most one score function for each vertex,
though we will allow multiple edges.
Then, up to constant factors the space required to specify an 
instance $F$ with constraint graph $G=(V,E,H)$
is the \emph{instance size} 
\begin{align}
 |F| &= 1+|V|+|E|+|H| .
\end{align}

We use the symbol \casesymbol~to end the
description of a reduction rule or the analysis of a case,
and \proofsymbol~to end a proof.

\section{Algorithm and outline of the analysis}  
 \label{algorithm}
 \label{outline}

We will show an algorithm (sketched as \alga) which,
on input of a hybrid instance~$F$, 
returns an optimal coloring $\phi$ of $F$'s vertices
in time 
$\Ostar{\twopow{w_e |E| + w_h |H|}}$,
which is to say in time 
\begin{align}
 T(F) &\leq \poly(|F|) \twopow{w_e |E| + w_h |H|}   \label{mainbound}
\end{align}
for some polynomial $\poly(\cdot)$.

\subsection{Algorithm and general arguments}
The algorithm is recursive:
on input of an instance~$F$,
in time polynomial in the instance size~$|F|$,
$F$ is \emph{reduced}
to a single instance $F'$ (a \emph{simplification}) 
or to several instances $F_1,\ldots,F_k$ (a \emph{splitting}), 
each of smaller size;
the algorithm solves the reduced instance(s) recursively;
and, again in time $\poly(|F|)$,
the algorithm constructs an optimal solution to $F$ from the
solutions of the reduced instances.

\begin{algorithm}[tbp]
\caption{Outline of algorithm and analysis}
\label{mainalg}
\begin{algorithmic}[1]
\STATE \textbf{Input:} 
 A hybrid \mts\ / 2-CSP instance $F$.
\STATE \textbf{Output:} 
 An optimal coloring $\phi$ of the vertices of~$F$.
\IF{$F$ has any vertex $v$ of degree $\geq 7$}
 \STATE Split on $\phi(v)=0$ and $\phi(v)=1$ to obtain $F_1$, $F_2$,
        recursively solve the instances $F_1$ and $F_2$ and return the
        best assignment for $F$.
 \STATE 
  \label{highdeg} 
  (Analysis:
  Inductively establish running time, using that both $F_1$ and $F_2$ have 
  at least $7$ edges fewer than $F$.)
\ENDIF
\STATE Simplify $F$.  (See Procedure~\ref{simplifyalg}.)
\STATE \label{algononsimplified}
 (Analysis:
 Establish running-time bound for general instances,
 using a bound for simplified instances.) 
\IF{$F$ is nonempty}
 \STATE Apply first applicable splitting reduction, obtaining $F_1,\ldots,F_k$.
 \STATE Simplify each of $F_1,\ldots,F_k$.
 \STATE Recursively solve $F_1,\ldots,F_k$ and return the
        best assignment for $F$.
 \STATE \label{algcore} 
   (Analysis: 
   Inductively establish running-time bound for simplified instances
   of maximum degree $\leq 6$,
   using $\sum_{i=1}^k \twopow{\mu(F_i)} \leq \twopow{\mu(F)}$.)
\ENDIF
\end{algorithmic}
\end{algorithm}

\begin{algorithm}[tbp]
\floatname{algorithm}{Procedure}
\caption{Simplification procedure}
\label{simplifyalg}
\begin{algorithmic}[1]
\STATE \textbf{Input:} 
 A hybrid instance $F$ 
\WHILE{Any of the following simplification rules is applicable}
 \STATE Apply the first applicable simplification:
 combine parallel edges;
 remove loops;
 0-reduction;
 delete a small component;
 delete a decomposable edge;
 half-edge reduction;
 1-reduction;
 1-cut;
 2-reduction;
 2-cut.
\ENDWHILE
\STATE Return the resulting simplified instance.
\end{algorithmic}
\end{algorithm}

The central argument
(corresponding to the analysis for line~\ref{algcore} of \alga)
is to establish \eqref{mainbound}
for \emph{simplified} formulas of maximum degree at most~$6$.
We do this shortly, in Lemma~\ref{mainlemma},
with the bulk of the paper devoted to verifying the lemma's hypotheses.

Given Lemma~\ref{mainlemma},
we then establish a similar running-time bound for instances
$F$ of degree at most $6$ which are not simplified,
that is, instances to which we may apply one or more of the simplifications
of \algb
(the analysis referred to by line~\ref{algononsimplified} in \alga),
and for instances of arbitrary degree
(the argument alluded to in line~\ref{highdeg} of \alga).

\subsection{Central argument}
The main argument is to establish \eqref{mainbound}
for simplified formulas of maximum degree at most $6$.
We will prove that
\begin{align}
 T(F) &= O(|F|^k \twopow{\mu(F)}) ,
\end{align}
which suffices if (as we will ensure)
for some constant $C$ and every 
simplified instance $F$ of degree at most $6$,
the \emph{measure} $\mu(F)$ satisfies
\begin{align}
 \mu(F) & \leq w_e |E| + w_h |H| + C . \label{munotmu}
\end{align}

In the following lemma's application, 
the class $\FF$ will consist of simplified hybrid formulas
of degree at most~6.

\begin{lemma}[Main Lemma] \label{mainlemma}
For a family $\FF$ of formulas,
suppose there exists an algorithm $A$ and a constant $c \ge 1$,
such that on input of any instance $F \in \FF$, 
$A$ either solves $F$ directly in time $O(1)$,
or decomposes $F$ into instances $F_1,\ldots,F_k \in \FF$,
solves these recursively, and \emph{inverts} their solutions to solve~$F$,
using time $O(|F|^c)$ for the decomposition and inversion
(but not the recursive solves). 
Further suppose that for a given measure~$\mu$,
\begin{align}
(\forall F \in \FF) \quad \mu(F) & \geq 0 , \label{mupos}
  \\
  \intertext{and, for any decomposition done by algorithm $A$,}
(\forall i) \quad |F_i| & \leq |F|-1 \text{, and}  \label{size}
  \\
\twomu(F_1) + \cdots + \twomu(F_k) & \leq \twomu(F) . \label{mu}
\end{align}
Then $A$ solves any instance $F \in \FF$ 
in time $O(|F|^{c+1}) \twomu(F)$.
\end{lemma}

We will often work with the equivalent to~\eqref{mu}, that
\begin{align}
 \sum_{i=1}^k \twopow{\mu(F_i)-\mu(F)} & \leq 1 . \label{mux} \tag{\ref{mu}$ '$}
\end{align}

\begin{proof}
The result follows easily by induction on~$|F|$.
Without loss of generality, we may replace the hypotheses' $O$ statements
with simple inequalities
(substitute a sufficiently large leading constant,
which then appears everywhere and has no relevance),
and likewise for the base case assume 
that we can return the solution $\emptyset$
to an empty instance in time~1.
If an instance $F$ is solved in time~1, then 
$T(F) = 1 \leq |F|^{c+1} \twomu(F)$.
Otherwise, where $T(F)$ denotes the time taken to solve an instance,
\begin{align*}
T(F) & \leq |F|^c + \ssum_{j=1}^k T(F_i) 
  &\text{\quad (by hypothesis)}
\\ &\leq |F|^c +\ssum |F_i|^{c+1} 2^{\mu(F_i)} 
  &\text{\quad (by the inductive hypothesis)}
\\ &\leq |F|^c + (|F|-1)^{c+1} \ssum 2^{\mu(F_i)}
  &\text{\quad (by hypothesis \eqref{size})}
\\ &\leq |F|^c + (|F|-1)^{c+1} 2^{\mu(F)} 
  &\text{\quad (by hypothesis \eqref{mu})}
\\ &\leq |F|^{c+1} 2^{\mu(F)} .
\end{align*}
The final inequality uses that $\mu(F) \geq 0$ and holds for any $c \geq 1$.
\end{proof}

The main work of the paper will be to find a 
set of decompositions and a measure $\mu$
such that the decompositions satisfy inequality~\eqref{size},
$\mu$ satisfies inequality~\eqref{mupos},
and (more interestingly)
$\mu$ satisfies inequality~\eqref{munotmu}
for some small values of $w_e$ and $w_h$,
and finally, for every decomposition, 
$\mu$ satisfies inequality~\eqref{mu}.

\subsection{Measure}
For  
an 
instance $F$ of (maximum) degree at most~6,
we define a measure $\mu(F)$ as
a sum of weights associated with 
light edges, heavy edges, 
and vertices of various degrees (at most~6),
and constants associated with the maximum degree $d$ of $F$
and whether $F$ is regular
(for all the degree criteria treating light and heavy edges alike):
\begin{align}
  \mu(F) & := \mua(F) + \mub(F) , \text{ with} \label{musum}
  \\
  \mua(F) & := |E| w_e + |H| w_h + \sum_{v \in V} w_{\deg(v)}  ,
 \\ 
  \mub(F) & := \murest .
	\label{measure}
\end{align}
Here $\chi(\cdot)$ is the indicator function:
1~if its argument is true, 0~otherwise.

To satisfy condition \eqref{munotmu}
it is sufficient that
\begin{align}
 (\forall d) \quad w_d \leq 0 ; \label{wneg}
\end{align}
this is also necessary for large regular instances.
Since we are now only considering instances of degree $\leq 6$,
we interpret ``$\forall d$'' to mean for all $d \in \set{0,1,\ldots,6}$.

\subsection{Peripheral arguments}

We first dispense with non-simplified instances.

\begin{lemma} \label{nonsimplifiedlemma}
Suppose that every \emph{simplified} \mc instance $F$ 
of degree at most $D \leq 6$
can be solved in time 
$\poly_1(|F|)  \twopow{\mu(F)}$. 
Suppose also that 
\begin{compactenum}
\item \label{dosimplify}
simplifying $F$
(or determining that $F$ is already simplified) 
takes time at most $\poly_2(|F|)$,
\item \label{measuresimplify}
any instance $F'$ 
obtained from simplifying $F$ satisfies
$|F'| \leq |F|-1$
and
$\mu(F') \leq \mu(F)+C'$
for some positive constant~$C'$,
and
\item \label{undosimplify}
the simplification can be \emph{reversed} in time at most $\poly_2(|F|)$
to recover an optimal solution to $F$
 from any optimal solution of $F'$.
\end{compactenum}
Then any instance $F$ of degree at most $D$ can be solved
in time
$\poly(|F|) \twopow{\mu(F)}$,
with $\poly(x) := \poly_2(x) + 2^{C'} \poly_1(x)$.
\end{lemma}

\begin{proof}
Since simplifying reduces the instance size,
a solution to the original instance $F$ can be obtained in time
\begin{align*}
T(F) &\leq \poly_2(|F|) + T(F')
 \\& \leq \poly_2(|F|) + \poly_1(|F'|) \twopow{\mu(F')}
 \\& \leq \poly_2(|F|) + \poly_1(|F|) \twopow{\mu(F)+C'}
 \\& \leq \big( \poly_2(|F|) + \twopow{C'} \poly_1(|F|) \big) \twopow{\mu(F)}
 \\& = \poly(|F|) \twopow{\mu(F)} .
\end{align*}
\end{proof}

The lemma's hypotheses \eqref{dosimplify} and \eqref{undosimplify}
will be satisfied by construction.
Hypothesis \eqref{measuresimplify} is assured if we constrain
that, for each simplification rule taking $F$ to $F'$,
\begin{align}
  \mua(F') \leq \mua(F) , \label{nonsimplified}
\end{align}
since by transitivity
the same inequality then holds for any sequence of simplifications 
starting with $F$ and ending with a simplified instance~$F'$,
and the desired inequality 
$\mu(F') = \mua(F)+\mub(F)-\mub(F') \leq \mua(F)+C'$
follows by boundedness of $\mub$ and choosing $C'$ sufficiently large.

Finally, we dispense with instances of high degree,
the argument alluded to in line~\ref{highdeg} of \alga.

\begin{lemma} \label{highdeglemma}
Suppose that every \mc instance $F$ of degree at most $6$
can be solved in time 
$O(|F|^{k_1} \twopow{w_e |E| + w_h |H|})$,
with $w_e, w_h \geq 1/7$.
Then for some sufficiently large~$k$,
every instance $F$ can be solved in time
$O(|F|^{k} \twopow{w_e |E| + w_h |H|})$.
\end{lemma}

\begin{proof}
As in the proof of Lemma \ref{mainlemma}, 
without loss of generality we may replace the $O$ statement
in the hypothesis with a simple inequality.
If $F$ has any vertex $v$ of degree at least $7$, we will set
$\phi(v)$ to 0 and 1 to generate instances $F_0$ and $F_1$ respectively, 
solve them recursively, and note that the solution to $F$
is that of the better of $F_0$ and~$F_1$, 
extended with the corresponding value for~$\phi(v)$.
We may assume that the splitting and its reversal together take time at most
$|F|^{k_2}$.

Ensure that $k \geq k_1$ is large enough that for all $x \geq 2$,
$x^{k_2} \leq x^k - (x-1)^k$,
and note that the hypothesis remains true replacing $k_1$ with~$k$.

The proof is by induction on~$F$.
If $F$ has no vertex of degree at least $7$ then we are already done.
Otherwise reduce $F$ to $F_1$ and $F_2$, 
each having at least $7$ fewer (light and/or heavy) edges than~$F$.
By induction we may assume the bound for $T(F_1)$ and $T(F_2)$,
so 
\begin{align*}
T(F) 
 &\leq |F|^{k_2}
   + 2 (|F|-1)^k \twopow{w_e |E| + w_h |H| - 7 \cdot 1/7}
 \\ &= |F|^{k_2} + (|F|-1)^k \twopow{w_e |E| + w_h |H|}
 \\ & \leq |F|^k \twopow{w_e |E| + w_h |H|} .
\end{align*}
The worst case for the last inequality 
is when $w_e |E| + w_h |H| =0$ (it is nonnegative),
and in that case the inequality follows by the construction of~$k$.
\end{proof}

\subsection{Optimizing the measure}
The task of the rest of the paper is to produce 
the comprehensive set of reductions hypothesized by Lemma~\ref{mainlemma}
(to any formula there should be some reduction we can apply)
and a measure $\mu$,
satisfying the hypotheses,
with $w_e$ as small as possible.
(More generally, if there are $m (1-p)$ conjunctions
and $mp$ general integer-valued clauses,
we wish to minimize $m(1-p)w_e + m p w_h$ or equivalently
$(1-p)w_e + p w_h$,
but for the discussion here we will just think in terms of
minimizing~$w_e$.)

For each reduction, 
the hypothesized constraint \eqref{size} will be trivially satisfied,
and it will be straightforward to write
down a constraint ensuring~\eqref{mux}.
We then solve the nonlinear program of minimizing $w_e$
subject to all the constraints.

Minimizing $w_e$ for a given set of constraints 
can be done with an off-the-shelf nonlinear solver
(see Section~\ref{solver}),
but finding a set of reductions resulting in a small value of $w_e$
remains an art.
It consists of trying some set of reductions, 
seeing which ones' constraints are tight in an optimal solution, 
and trying to replace these reductions with more favorable ones.

\newcommand{\obj}{w}

\begin{table}
 \begin{tabular}{||c|ccc|ccc|ccc||}
 \hline \hline
  \multicolumn{1}{||r|}{$p$}
    & & 0 &                       & & 0.05 &                    & & 0.1 &\\
  $\Delta(F)$
    &$w_e$ &$w_h$ &$\obj$&$w_e$ &$w_h$ &$\obj$&$w_e$ &$w_h$ &$\obj$\\
  \hline
      3 & 0.10209 & 0.23127 & 0.10209 & 0.10209 & 0.23125 & 0.10855 & 0.10209 & 0.23125 & 0.11501\\
      4 & 0.14662 & 0.31270 & 0.14662 & 0.14662 & 0.31270 & 0.15493 & 0.15023 & 0.26951 & 0.16216\\
      5 & 0.15518 & 0.30728 & 0.15518 & 0.15637 & 0.27997 & 0.16255 & 0.15640 & 0.27951 & 0.16871\\
$\ge$ 6 & 0.15819 & 0.31029 & 0.15819 & 0.15912 & 0.28223 & 0.16527 & 0.15912 & 0.28223 & 0.17143\\
 \hline \hline
 \end{tabular}
 \medskip
 
 \begin{tabular}{||c|ccc|ccc|ccc||}
 \hline \hline
  \multicolumn{1}{||r|}{$p$}
    & & 0.2 &                     & & 0.3 &                     & & 1 &\\
  $\Delta(F)$
    &$w_e$ &$w_h$ &$\obj$&$w_e$ &$w_h$ &$\obj$&$w_e$ &$w_h$ &$\obj$\\
  \hline
      3 & 0.10209 & 0.23125 & 0.12793 & 0.10209 & 0.23125 & 0.14084 & 0.16667 & 0.16667 & 0.16667\\
      4 & 0.15023 & 0.26951 & 0.17409 & 0.15023 & 0.26951 & 0.18601 & 0.18750 & 0.18750 & 0.18750\\
      5 & 0.15640 & 0.27951 & 0.18102 & 0.19000 & 0.19000 & 0.19000 & 0.19000 & 0.19000 & 0.19000\\
$\ge$ 6 & 0.16520 & 0.25074 & 0.18231 & 0.19000 & 0.19000 & 0.19000 & 0.19000 & 0.19000 & 0.19000\\
 \hline \hline
 \end{tabular}
 \medskip
 
 \caption{Values of $w_e$, $w_h$ and $\obj := p w_h+(1-p)w_e$ 
  according to the fraction $p$
  of heavy edges and the maximum degree $\Delta(F)$ of a formula $F$.
  For any pair $(w_e,w_h)$ in the table, 
  a running-time bound of $\Ostar{\twopow{m \cdot ((1-p)w_e+p w_h)}}$
  is valid for every formula, 
  regardless of its fraction $p(F)$ of non-simple clauses,
  but the pair obtained when the table's $p$ equals $p(F)$ 
  gives the best bound.
  \label{tab:runtimes}
 }
\end{table}

With the constraints established in the next sections, 
we will obtain our main result.

\begin{theorem} \label{thm:runtimes}
Let $F$ be an instance of integer-weighted \mc
in which each variable appears in at most $\Delta(F)$ 2-clauses,
and there are $(1-p(F))m$ conjunctive and disjunctive 2-clauses,
and $p(F) m$ other 2-clauses. 
Then, for any pair of values $w_e, w_h$ in Table~\ref{tab:runtimes}
(not necessarily with the table's $p$ equal to $p(F)$),
the above algorithm solves $F$ in time
$\Ostar{\twopow{m \cdot ((1-p(F))w_e+p(F) w_h)}}$.
When the table's $p = p(F)$, we obtain our best bound,
$\Ostar{\twopow{m \cdot ((1-p)w_e+p w_h)}} =
 \Ostar{\twopow{m w}}$.
\end{theorem}

\begin{proof}
Corollary of Lemma~\ref{mainlemma}, 
solving the mathematical program given by the various constraints
given in the next sections
and minimizing~$p w_h+(1-p)w_e$.
\end{proof}

Which of the constraints are tight strongly depends on $p$ and $\Delta(F)$.

\subsection{The measure's form} \label{measuresform}
Let us explain the rather strange form of the measure.
Ideally, it would be defined simply as~$\mua$,
and indeed for the measure we ultimately derive,
all of our simplifications and most of our splittings 
satisfy the key inequality \eqref{mux}
with $\mua$ alone in place of $\mu$.
Unfortunately, for regular instances of degrees 4, 5, and~6,
satisfying this constraint would require a larger value of~$w_e$.
Viewing \eqref{mux} equivalently as
\begin{align*}
 \sum_{i=1}^k \twopow{\mu(F_i)-\mu(F)} & \leq 1 , 
\end{align*}
adding a cost $R_d$ to the measure of a $d$-regular instance $F$
means that if a $d$-regular instance $F$ 
is reduced to nonregular instances $F_1$ and~$F_2$
of degree $d$,
each difference $\mu(F_i)-\mu(F)$ is smaller by $R_d$
than the corresponding difference $\mua(F_i)-\mua(F)$.
We will therefore want
\begin{align} \label{Rpos}
  (\forall d \in \set{4,5,6}) \quad R_d \geq 0 .
\end{align}
Of course, if a nonregular instance $F$ of degree $d$ is reduced to
instances $F_i$ of degree $d$ one or more of which is regular,
there will be a corresponding penalty:
for each $d$-regular~$F_i$, 
$\mu(F_i)-\mu(F)$ is $\mua(F_i)-\mua(F)+R_d$.

Indeed, for each splitting reduction we will have to consider 
several cases. 
Typically, the ``baseline'' case will be the reduction of a 
nonregular instance to two nonregular instances.
In this case $\mu$ and $\mua$ are equivalent,
and if we know for example that $\mua(F_i)-\mua(F) \leq x_i$, 
our nonlinear program constrains that $\twopow{x_1} + \twopow{x_2} \leq 1$.

If we reduce starting from a regular instance,
the nature of the reductions is such that, generically,
we will get less favorable bounds 
$\mua(F_i)-\mua(F) \leq x'_i$ 
(the values $x'_i$ will be larger than the $x_i$ were),
but we also get a ``reward'' (a further decrease of~$R_d$) 
for whichever of $F_1$ and $F_2$ are not also regular.
If we reduce starting from a nonregular instance 
but producing one or more regular children,
we will consider various possibilities.

The case where we a nonregular instance of degree $d$
produced a regular instance $F_i$ of degree $<d$,
can be dispensed with simply by choosing $C_d$ sufficiently large,
to reap whatever additional reward is needed.
Our splitting rules are generally local and will never increase 
measure by more than a constant, so some constant $C_d$ suffices.
Also, our reductions never increase the degree of an instance
(each $F_i$ has degree at most that of~$F$),
so $C_d$ will never work against us,
and there is no harm in choosing it as large as we like. 
Thus, we never need to consider the particulars of cases
where the instance degree decreases,
nor the values~$C_d$.

The remaining cases where a nonregular instance has regular children
will be considered on a case-by-case basis for each reduction. 
Generally, for a child to become regular means that,
beyond the constraint-graph changes taken into account 
in the baseline case (with the child nonregular),
some additional vertices (those of degree less than~$d$)
must have been removed from the instance by simplifications.
Accounting for these implies a further decrease in measure 
that compensates for the increase by~$R_d$.

\section{Some initial constraints}  \label{sec:measure}

We have already derived one constraint for $\mu$, namely~\eqref{wneg},
and we will now 
introduce some notation and
derive several more constraints.

Let us write $w(v)$ for the weight of a vertex~$v$
(so $w(v)=w_d$ for a vertex of degree~$d$), 
and similarly $w(e)$ for the weight of an edge
($w_e$ or $w_h$ depending on whether $e$ is light or heavy).
Sometimes it will be helpful to think of $\mua(F)$ as
\begin{align}
  \mua(F) &= 
   \sum_{v \in V} \Big( w(v) + \tfrac12 \sum_{e \colon v \in e} w(e) \Big) ,
	\label{measure2}
\end{align}
the sum of the weights of the vertices and their incident \emph{half edges}.
For convenience, we define (and thus constrain)
\begin{align}
	\constrain{ a_d  = w_d + \tfrac12 d w_e } .
	\label{ai}
\end{align}
Thus, $a_d$ is equal to the summand in \eqref{measure2}
for a vertex all of whose incident edges are light, and smaller
otherwise.

We require $\mu(F) \geq 0$ for all instances.
Considering regular \mts instances with
degree~$d$ ($d=0,\ldots,6$), this implies that
\begin{align}
  \constrain{ (\forall d) \quad a_d \geq 0 } . \label{apos} 
\end{align}
(For $d \leq 3$, \eqref{apos} is implied by
$\mub(F)=0$, with \eqref{measure2} and~\eqref{ai}.
For $d \geq 4$, positivity of $\mua$ 
might give positive measure to $K_d$
even if $\mub(K_d)$ were negative, but then a graph consisting of 
sufficiently many copies of $K_d$ would still have negative measure.)
If we also constrain that
\begin{align}
  \constrain{ (\forall d \in \set{4,5,6}) \quad C_d, R_d \geq 0 } ,
     \label{constpos}
\end{align}
then we have assured that $\mu(F) \geq 0$ for all instances.
In the end, constraint \eqref{constpos} will not be tight and so there is 
no loss in making the assumption.

Were it the case that $w_h \leq w_e$, then we could simply transform each
light edge into a heavy one, reducing the measure,
and getting a better time bound for solving 
an instance of \mc than an instance of \mts or a hybrid instance.
Thus if we are to gain any advantage from considering \mts 
or hybrid instances, it must be that
\begin{align}
  \constrain{ w_e \leq w_h . } \label{whwe}
\end{align}
In the end we will find that this constraint is not tight, 
and so there is no cost to making the assumption.%
\footnote{%
For the most part we will only write down constraints that are 
\emph{necessary}, typically being required for some reduction 
to satisfy~\eqref{mux}, but we make a few exceptions early on.}%

For intuitive purposes let us leap ahead and 
mention that we will find that $a_0 = a_1  = a_2 = 0$,
(thus $w_0=0$, $w_1 = -\tfrac12 w_e$, and $w_2 = -w_e$),
while $0 < a_3 < \cdots < a_6$.
Per \eqref{whwe} above, $w_h \geq w_e$.
Typically we will find that $w_h \leq 2 w_e$,
but not always.
(Even where this fails to hold, notably for cubic \mts,
we can still replace two conjunctions or disjunctions on the same variables
with one CSP edge:
decreasing the degrees of the incident vertices 
decreases the measure enough to make up for the increase of $w_h-2w_e$.)
This ``intuition'' has changed several times as the paper has evolved,
which supports the value of making as few assumptions as possible,
instead just writing down constraints implied by the reductions.

\section{Simplification rules and their weight constraints}
\label{simplification}

We use a number of simplification rules
(reductions of $F$ to a single simpler instance $F_1$ or~$F'$).
Some of the simplification rules are standard, 
the CSP 1-reductions are taken from~\cite{faster},
the CSP 2-reductions combine ideas from \cite{faster} and~\cite{KK},
and a ``super 2-reduction'' is introduced here.
For vertices of degree 5
we use a splitting reduction taken from~\cite{clauseLearning}
that we generalize to hybrid instances.

We have already ensured 
constraint \eqref{mupos} by \eqref{apos} and \eqref{constpos},
so our focus is on ensuring that each reduction satisfies~\eqref{mux}.
Since each splitting is followed by an (unpredictable)
sequence of simplifications, 
to have any hope of satisfying \eqref{mux} it is essential
that each simplification from any $F$ to $F'$ satisfies
\begin{align}
  \mua(F') &\leq \mua(F) ; \label{X}
\end{align}
in any case this inequality is required by
Lemma \ref{nonsimplifiedlemma} 
(it duplicates inequality~\eqref{nonsimplified}).
Constraint \eqref{size} of Lemma~\ref{mainlemma}
will be trivially satisfied by all our simplifications
and splittings.

Recapitulating, in this section we show that \eqref{X}
is satisfied by all our simplifications. 
Ensuring \eqref{mux} will come when we look at the splitting rules,
and the measure component $\mub$ we are ignoring here.

\myitem{Combine parallel edges} \label{combineparallel}
Two parallel edges (light or heavy) with endpoints $x$ and $y$
may be collapsed into a single heavy edge.
This means that the ``transformed'' instance $F'$
($F_1$ in Lemma~\ref{mainlemma}, with $k=1$)
is identical to $F$ except that the two score functions 
$s_{xy}(\phi(x),\phi(y))$
and
$s'_{xy}(\phi(x),\phi(y))$
in $F$ are replaced by their sum
$s''_{xy}(\phi(x),\phi(y))$ in $F'$.
If one of the endpoints, say $x$, of the two parallel edges
has degree~$2$, collapse the parallel edges and immediately
apply a 1-reduction (see \ref{1reduce}) on $x$
(of degree~1), which removes $x$ from the constraint graph.
To ensure \eqref{X} we constrain
\begin{align}
  (\forall d \geq 2) \quad - a_2 - a_d + a_{d-2} \leq 0 : \label{heavy2}
\end{align}
the left hand side is $\mua(F')-\mua(F)$ 
thought of as the subtraction of a vertex of degree~2, a vertex
of degree~$d$ and the addition of a vertex of degree~$d-2$.
For the case that $x$ and $y$ have degree $d \ge 3$,
we constrain
\begin{align}
  (\forall d \geq 3) \quad - 2 a_d + 2 a_{d-1} - w_e + w_h  \leq 0 : \label{heavy}
\end{align}
the left hand side is $\mua(F')-\mua(F)$
thought of as replacing two vertices of degree $d$ by two vertices
of degree $d-1$ and replacing a light edge by a heavy edge.
(Remember that the degree of a vertex is the number 
of incident edges rather than
the number of distinct neighbors.)
If $\deg(x) \neq \deg(y)$, the resulting constraint is a 
half--half mixture of a constraint~\eqref{heavy} 
with $d=\deg(x)$ and another with $d=\deg(y)$,
and is thus redundant.

By construction, the score functions of $F'$ and $F$ are identical,
so an optimal solution $\phi'$ for $F'$ \emph{is}
an optimal solution $\phi$ of $F'$ (no transformation is needed).
\caseend

Applying this reduction whenever possible, 
we may assume that the instance has no parallel edges.

Note that we cannot hope to combine simple clauses (conjunctions and
disjunctions) and still take 
advantage of their being simple clauses rather than general CSP clauses: 
$(x \OR y)+(\bar x \OR \bar y) = 1+(x \XOR y)$,
the additive 1 is irrelevant, and the XOR funtion is not simple.

\myitem{Remove loops} 
If the instance includes any edge $xx \in E \cup H$, 
the nominally dyadic score function $s_{xx}(\phi(x),\phi(x))$ 
may be replaced by a (or incorporated into an existing) 
monadic score function $s_x(\phi(x))$.
This imposes the constraints
\begin{align}
  (\forall d \geq 2) \quad - a_d + a_{d-2} \leq 0. \label{loops}
\end{align}
\caseend

As this constraint is stronger than \eqref{heavy2}, we may
ignore constraint \eqref{heavy2}.

With this and the edge-combining reduction, we may 
at all times assume the constraint graph is simple.

\myitem{Delete a vertex of degree 0 (0-reduction)} 
If $v$ is a vertex of degree~0,
reduce the instance $F$ to $F'$ 
by deleting $v$ and its monadic score function~$s_v$,
solve $F'$, and obtain an optimal solution of $F$ by
augmenting the solution of $F'$
with whichever coloring $\phi(v)$ of $v$ 
gives a larger value of $s_v(\phi(v))$.
Constraint~\eqref{size} is satisfied, since 
$|F'|=|F|-1$.
Constraint \eqref{X} is satisfied if and only if
$ -w_0 \leq 0 $.
On the other hand, for a useful result we need each $w_d \leq 0$
(inequality~\eqref{wneg}), implying that $w_0=0$, and thus 
\begin{align}
 a_0 = 0 . \label{a00}
\end{align}
We will henceforth ignore vertices of degree 0 completely.
\caseend

\myitem{Delete a small component}  \label{smallcomponent}
For a constant $C$ (whose value we will fix
in the splitting reduction \myref{largecomponent}),
if the constraint graph $G$ of $F$ has components $G'$ and~$G''$
with $1 \leq |V(G'')| < C$
($|V(G')|$ is arbitrary),
then $F$ may be reduced to $F'$ with constraint graph~$G'$.
The reduction and its correctness are obvious,
noting that $F''$ may be solved in constant time.
Since $\mua(F')-\mua(F) \leq -\sum_{v \in V(G)} a_{\deg(v)}$,
it is immediate from \eqref{apos} that \eqref{X} is satisfied.
\caseend

\myitem{Delete a decomposable edge} \label{decomposablered}
If a dyadic score function $s_{xy}(\phi(x),\phi(y))$ 
can be expressed as a sum of monadic scores,
$s'_x(\phi(x))+s'_y(\phi(y))$,
then delete the edge and
add $s'_x$ to the original $s_x$, and $s'_y$ to $s_y$.
If $x$ and $y$ have equal degrees, the constraint imposed is that 
$(\forall d \geq 1)$ $-w_e - 2 w_d + 2 w_{d-1} \leq 0$,
or equivalently,
\begin{align}
(\forall d \geq 1) \quad  - a_d +  a_{d-1} \leq 0 .
 \label{decomposable}
\end{align}
(The $d=1$ case was already implied by 
\eqref{a00} and \eqref{apos}.)
As in \eqref{heavy}, inequalities for degree pairs
are a mixture of those for single degrees.
Note that we may ignore constraint \eqref{loops} now as it is weaker
than \eqref{decomposable}.
\caseend

Three remarks.
First, together with~\eqref{a00}, 
\eqref{decomposable} means that
\begin{align}
 0 = a_0 \leq a_1 \leq \cdots \leq a_6 . \label{aincreasing}
\end{align}

Second, if an edge is not decomposable,
the assignment of either endpoint has a (nonzero) bearing on 
the optimal assignment of the other, 
as we make precise in Remark~\ref{nondecomposable}.
We will exploit this in Lemma~\ref{super2gen}, which shows how 
``super 2-reduction'' opportunities \myref{super2} are created.
\begin{remark} \label{nondecomposable}
Let 
 \begin{align*}
 \bias_y(i) &:= s_{xy}(i,1) - s_{xy}(i,0) ,
 \end{align*}
the ``preference'' of the edge function $s_{xy}$ 
for setting $\phi(y)=1$ over $\phi(y)=0$
when $x$ is assigned $\phi(x)=i$.
Then $s_{xy}$ is decomposable if and only if $\bias_y(0) = \bias_y(1)$.
\end{remark}
\newcommand{\sxy}{s_{xy}}
\begin{proof}
We have that $s_{xy}$ is decomposable if and only if its 
2-by-2 table of function values has rank~1,
which is equivalent to equality of the two diagonal sums,
$\sxy(0,1)+\sxy(1,0) = \sxy(0,0)+\sxy(1,1)$, 
which in turn is equivalent to 
$\sxy(0,1)-\sxy(0,0) = \sxy(1,1)-\sxy(1,0)$,
i.e.,
$\bias_y(0) = \bias_y(1)$.
\end{proof}

Finally, 
when some vertices and their incident edges are deleted from a graph,
we may think of this as the deletion of each vertex
and its incident half-edges
(which typically we will account for explicitly)
followed (which we may not account for)
by the deletion of any remaining half-edges 
and the concomitant decrease in the degrees of their incident vertices
(for edges one of whose endpoints was deleted and one not).
A ``half-edge deletion'' and vertex degree decrease
is precisely what is characterized
by the left-hand side of \eqref{decomposable},
so it cannot increase the measure~$\mua$.
Even though such simplifications take place on an intermediate structure 
that is more general than a graph, and that we will not formalize,
for convenient reference we will call this a half-edge reduction.

\myitem{Half-edge reduction} \label{halfedge}
Delete a half-edge, and decrease the degree of its incident vertex.
By~\eqref{decomposable}, this does not increase the measure.

\myitem{Delete a vertex of degree 1 (1-reduction)} \label{1reduce}
This reduction comes from~\cite{faster},
and works regardless of the weight of the incident edge.
Let $y$ be a vertex of degree~1, with neighbor~$x$.
Roughly, we use the fact that the optimal assignment of $y$ 
is some easily-computable function of the assignment of~$x$, 
and thus $y$ and its attendant score functions $s_y(\phi(y))$
and $s_{xy}(\phi(x),\phi(y))$ can be incorporated into 
$s_x(\phi(x))$.

\newcommand{\monad}[3]{{#3}_{#1}({#2})}
\newcommand{\s}[2]{\monad{#1}{#2}{s}}
\renewcommand{\sp}[2]{\monad{#1}{#2}{s'}}
\newcommand{\dyad}[5]{{#5}_{#1 #2}({#3 #4})}
\renewcommand{\ss}[4]{\dyad{#1}{#2}{#3}{#4}{s}}
\newcommand{\ssp}[4]{\dyad{#1}{#2}{#3}{#4}{s'}}
\newcommand{\snought}{s_\emptyset}

We take a precise formulation from~\cite{faster}.
Here $V$ is the vertex set of $F$, $E$~is the set of all edges
(light and heavy), and $S$ is the set of score functions.

Reducing $(V,E,S)$ on $y$ results in a new instance $(V',E',S')$
with $V' = V \setminus y$ and $E' = E \setminus xy$.
$S'$ is the restriction of $S$ to $V'$ and $E'$, except that
for all ``colors'' $C \in \bool$
we set
\begin{align*}
\sp xC & = \s xC + \max_{D \in \bool} \{\ss xyCD + \s yD\} .
\end{align*}

Note that any coloring $\phi'$ of $V'$ can be extended to
a coloring $\phi$ of $V$ in two ways, depending on the color assigned to~$y$.
Writing $(\phi',D)$ for the extension in which $\phi(y)=D$,
the defining property of the reduction is that
$S'(\phi') = \max_D S(\phi',D)$.
In particular,
$\max_{\phi'}S'(\phi') = \max_{\phi}S(\phi)$,
and an optimal coloring $\phi'$ for the instance $(V',E',S')$
can be extended to an optimal coloring $\phi$ for $(V,E,S)$.
This establishes the validity of the reduction.

Since the reduction deletes the vertex of degree~1 and its incident
edge (light, in the worst case), and decreases the degree of 
the adjacent vertex,
to ensure \eqref{X}, we constrain that
 $(\forall d \geq 1)$ $-w_1 -w_e -w_d + w_{d-1} \leq 0$,
or equivalently that
\begin{align*}
 (\forall d \geq 1) \quad  a_{d-1} - a_d - a_1 \leq 0 ,
\end{align*}
which is already ensured by \eqref{aincreasing}.
\caseend

\myitem{1-cut} \label{1cut}
Let $x$ be a cut vertex isolating a set of vertices~$A$,
$2 \leq |A| \leq 10$.
(The 1-cut reduction extends the 1-reduction, thought of as the case $|A|=1$.)
Informally, for each of $\phi(x) = 0,1$
we may determine the optimal assignments of the vertices in $A$ 
and the corresponding optimal score; 
adding this score function to the original monadic score $s_x$ 
gives an equivalent instance $F'$ on variables $V \setminus A$.
With $A$ of bounded size, construction of~$F'$, and
extension of an optimal solution of $F'$ to one of~$F$, 
can be done in polynomial time.
(Formal treatment of a more general ``cut reduction''
on more general ``Polynomial CSPs'' can be found in~\cite{countingArxiv3}.)

This simplification imposes no new constraint on the weights.
Vertices in $A$ and their incident half-edges are deleted, 
and any remaining half-edges (those incident on~$x$)
are removed by half-edge reductions~\myref{halfedge};
by \eqref{aincreasing},
neither increases the measure~$\mua$.
\caseend

\myitem{Contract a vertex of degree 2 (2-reduction)} \label{2reduce}
Let $y$ be a vertex of degree 2 with neighbors $x$ and~$z$.
Then $y$ may be contracted out of the instance: 
the old edges $xy$, $yz$, and (if any) $xz$
are replaced by a single new edge $xz$ which in general is heavy,
but is light if there was no existing edge $xz$
and at least one of $xy$ and $yz$ was light.

The basics are simple, but care is needed both because of
the distinction between light and heavy edges
and because we insist that the constraint
graph be simple, and the 2-reduction is the one operation that 
has the capacity to (temporarily) create parallel edges
and in the process change the vertex degrees.
We consider two cases: 
there is an edge $xz$; and
there is no edge~$xz$.

If there is an edge $xz$ then $x$ and $z$ both have degree 3 or more by Simplification~\ref{1cut}, 
we use the general \mc 2-reduction from~\cite{faster}.  
Arguing as in the 1-reduction above, 
here the optimal assignment of $y$
depends only on the assignments of $x$ and~$z$,
and thus we may incorporate all the score terms involving~$y$, namely
$s_y(\phi(y)) + s_{xy}(\phi(x),\phi(y)) +s_{yz}(\phi(y),\phi(z))$,
into a new $s'_{xz}(\phi(x),\phi(z))$,
which is then combined with the original $s_{xz}(\phi(x),\phi(z))$.
The effect is that $y$ is deleted, 
three edges (in the worst case all light) are replaced by one heavy edge,
and the degrees of $x$ and $z$ decrease by one.
If $\deg(x)=\deg(y)=d$, $\mua(F')-\mua(F) \leq 0$ 
is assured by
$ -w_2 -3 w_e + w_h - 2 w_d + 2 w_{d-1} \leq 0$,
or equivalently
\begin{align*}
(\forall d \geq 3) \quad  
 -a_2 - w_e + w_h - 2 a_d + 2 a_{d-1} \leq 0 ,
\end{align*}
which is already ensured by \eqref{apos} and \eqref{heavy}.
As in \eqref{decomposable}, inequalities for pairs $\deg(x) \neq \deg(y)$
are a mixture of those for single degrees.
\emph{If $xy$ or $yz$ is heavy, then $\mua(F')-\mua(F) \leq -w_h + w_e$,
and we will capitalize on this later.}

Finally, we consider the case where there was no edge~$xz$.
If $xy$ and $yz$ are both heavy, then as in the first case we 
apply the general \mc reduction to replace them with a
heavy edge~$xz$,
giving 
$\mua(F')-\mua(F) 
 \leq -2w_h + w_h - w_2 
 = -a_2 -w_h + w_e 
 \leq -w_h + w_e$.

Otherwise, at least one of $xy$ and $yz$ is light,
and we show that the resulting edge $xz$ is light.
(For pure Sat formulas, this is the ``frequently meeting variables'' rule 
of~\cite{KK}.)
Without loss of generality we assume that $xy$ 
is the conjunctive constraint $x \OR y$
or the disjunction $x \AND y$
(what is relevant is that the clause's score is restricted
to $\set{0,1}$, and is monotone in~$\phi(y)$).
We define a \emph{bias} 
\begin{align}
\bias_y(i) &= [s_y(1)-s_y(0)] + [s_{yz}(1,i)-s_{yz}(0,i)] ,
\end{align}
to be the ``preference'' (possibly negative)
of $s_y + s_{yz}$ for setting $\phi(y)=1$ versus $\phi(y)=0$,
when $z$ has been assigned $\phi(z)=i$.
If $\bias_y(i) \leq -1$ then $\phi(y)=0$
is an optimal assignment.
(That is, for every assignment to the remaining variables,
including the possibility that $\phi(x)=0$,
setting $\phi(y)=0$ yields at least as large as score as $\phi(y)=1$.)
Also, if $\bias_y(i) \geq 0$ then $\phi(y)=1$ is an optimal assignment. 

Thus, an optimal assignment $\phi(y)$ can be determined 
as a function of $\phi(z)$ alone,
with no dependence on $\phi(x)$.
(This cannot be done in the general case
where $xy$ and $yz$ are both heavy edges.)
With $\phi(y)$ a function of $\phi(z)$,
the score $s_{yz}(\phi(y),\phi(z))$ 
may be incorporated into the monadic score function~$s_z(\phi(z))$.
Also, there are only 4 functions from $\bool$ to $\bool$:
as a function of $\phi(z)$,
$\phi(y)$ must the constant function~0 or 1
(in which cases $x \OR y$ can be replaced respectively
by a monadic or niladic clause)
or $\phi(z)$ or $\overline{\phi(z)}$
(in which cases $x \OR y$ can be replaced respectively 
by the Sat clause $x \OR z$ or $x \OR \bar{z}$).

This shows that if there is no edge $xz$ and
either $xy$ or $yz$ is light,
then the 2-reduction produces a light edge~$xz$.
If both $xy$ and $yz$ are light,
$\mua(F')-\mua(F) 
 \leq -a_2 \leq 0$,
while (once again) if one of $xy$ and $yz$ is heavy,
$\mua(F')-\mua(F) \leq -w_h +w_e$.

To summarize, 
no new constraint is imposed by 2-reductions. 
Also, if either of $xy$ or $yz$ is heavy,
then we have not merely that $\mua(F')-\mua(F) \leq 0$
but that $\mua(F')-\mua(F) \leq -w_h+w_e$,
and we will take advantage of this later on.
\caseend

\myitem{2-cut} \label{2cut}
Let $\set{x,y}$ be a 2-cut isolating a set of vertices~$A$,
$2 \leq |A| \leq 10$.
(The 2-cut reduction extends the 2-reduction, 
thought of as the case $|A|=1$.)
Similarly to the 1-cut above,
for each of the four cases $\phi: \set{x,y} \to 0,1$
we may determine the optimal assignments of the vertices in $A$ 
and the corresponding optimal score; 
adding this score function to the original dyadic score $s_{xy}$ 
gives an equivalent instance $F'$ on variables $V \setminus A$.
There is nothing new in the technicalities, and we omit them.

In general, $\mua'-\mua$ may be equated with the weight change from 
deleting the original edge $xy$ if any
(guaranteed by \eqref{decomposable} not to increase the measure),
deleting all vertices in $A$ with their incident half edges
(a change of $-\sum_{v \in A} a_{\deg(v)}$),
replacing one half-edge from each of $x$ and $y$ into $A$
with a single heavy edge between $x$ and $y$ 
(not affecting their degrees, and thus a change of $-w_e+w_h$),
then doing half-edge reductions to remove any half-edges remaining
 from other edges in $\set{x,y} \times A$ 
(guaranteed by reduction \ref{halfedge} not to increase the measure).
Thus, 
$-\sum_{v \in A} a_{\deg(v)} -w_e+w_h \leq -2 a_3 -w_e+w_h$,
where the second inequality uses that
$|A|\geq 2$, all vertices have degree $\geq 3$
(a 2-reduction is preferred to this 2-cut reduction),
and the values $a_i$ are nondecreasing (see~\eqref{aincreasing}).
Thus we can assure that $\mua'-\mua \leq 0$ by
\begin{align*}
 -2 a_3 - w_e + w_h &\leq 0 ,
\end{align*}
which is already imposed by \eqref{apos} and \eqref{heavy}.
\caseend

\section{Some useful tools}

Before getting down to business, 
we remark that in treating disjunction and conjunction efficiently,
as well as decomposable functions 
(see reduction \ref{decomposablered} and Remark~\ref{nondecomposable}),
the only boolean function our algorithm cannot treat efficiently
is exclusive-or. 
The following remark is surely well known.

\begin{remark} \label{booleanfunctions}
The only non-decomposable two-variable boolean functions
are conjunction, disjunction, and exclusive-or.
\end{remark}
\begin{proof}
A function $s \colon \bool^2 \mapsto \set{0,1}$
is characterized by a $2 \times 2$ table of 0s and 1s.
If the table has rank~1 (or~0), we can decompose $s$ into 
monadic functions 
writing $s_{xy}(\phi(x),\phi(y)) = s_x(\phi(x)) + s_y(\phi(y))$.
A table with zero or four 1s is a constant function, trivially decomposable.
A table with one 1 is the function $\phi(x) \AND \phi(y)$,
up to symmetries of the table and (correspondingly)
negations of one or both variables;
similarly a table with three 1s is the function $\phi(x) \OR \phi(y)$.
In a table with two 1s, either the 1s share a row or column,
in which case the function is decomposable,
or they lie on a diagonal, which is (up to symmetries and signs)
the function $\phi(x) \oplus \phi(y)$.
\end{proof}
The property of disjunction and conjunction on which we rely
(besides having range $\set{0,1}$)
is that they are monotone in each variable. 
Obviously exclusive-or is not monotone,
and it seems that it cannot be accommodated by our methods.

\myitem{Super 2-reduction} \label{super2}
Suppose that $y$ is of degree~2 
and that its optimal color {$C \in \bool$} is independent of 
the colorings of its neighbors $x$ and~$z$, i.e., 
\begin{align}
  (\forall D,E) \quad 
    & s_y(C) + s_{yx}(C,D) + s_{yz}(C,E)
   \\ &=  \quad \nonumber
    \max_{C' \in \bool} s_y(C') + s_{yx}(C',D) + s_{yz}(C',E) .
\end{align}

In that case,
$s_y(\phi(y))$ can be replaced by $s_y(C)$ and incorporated 
into the niladic score,
$s_{xy}(\phi(x),\phi(y))$ can be replaced by a monadic score
$s'_x(\phi(x)) := s_{xy}(\phi(x),C)$ 
and combined with the existing $s_x$,
and the same holds for $s_{yz}$, 
resulting in an instance with $y$ and its incident edges deleted.
\caseend

A super 2-reduction is better than a usual one
since $y$ is deleted, not just contracted.

We will commonly \emph{split} on a vertex $u$,
setting $\phi(u)=0$ and $\phi(u)=1$ to obtain instances $F_0$ and $F_1$,
and solving both.

\begin{lemma} \label{super2gen}
After splitting a simplified instance $F$
on a vertex $u$ incident to a vertex $y$ of degree~3
whose other two incident edges $xy$ and $yz$ are both light, 
in at least one of the reduced instances $F_0$ or~$F_1$,
$y$ is subject to a super 2-reduction.
\end{lemma}

\begin{proof}
In the clauses represented by the light edges $xy$ and $yz$,
let $b \in \set{-2,0,2}$ 
be the number of occurrences of $y$ 
minus the number of occurrences of~$\bar y$.
(As in reduction~\ref{2reduce},
we capitalize on the fact that conjunction and disjunction
are both elementwise monotone, 
and that their scores are limited to $\set{0,1}$.)
Following the fixing of~$u$ to 0 or~1 and its elimination,
let $\bias_y := s_y(1)-s_y(0)$.
Given that $F$ was simplified,
the edge $uy$ was not decomposable,
so by Remark~\ref{nondecomposable}
the value of $\bias_y$ in $F_0$
is unequal to its value in $F_1$.

First consider the case $b=0$.
If $\bias_y \geq 1$, the advantage from $\bias_y$
for setting $\phi(y)=1$ rather than 0
is at least equal to the potential loss (at most~1)
 from the one negative occurrence of $y$ in $xy$ and $yz$,
so the assignment $\phi(y)=1$ is always optimal.
Symmetrically, if $\bias_y \leq -1$ we may set $\phi(y)=0$.
The only case where we cannot assign $y$ is when $\bias_y = 0 = -b/2$.

Next consider $b=2$.
(The case $b=-2$ is symmetric.)
If $\bias_y \geq 0$ we can fix $\phi(y)=1$,
while if $\bias_y \leq -2$ we can fix $\phi(y)=0$.
The only case where we cannot assign $y$ is when $\bias_y = -1 = -b/2$.

Thus, we may optimally assign $y$ 
independent of the assignments of $x$ and $z$
unless $\bias_y = -b/2$.
Since $\bias_y$ has different values in $F_0$ and~$F_1$,
in at least one case $\bias_y \neq -b/2$
and we may super 2-reduce on~$y$.
\end{proof}

\subsection{Splitting on vertices of degree 5}
Kulikov and Kutzkov~\cite{clauseLearning} introduced a clever 
splitting on vertices of degree~5. 
Although we will not use it until we address instances
of degree~5 in Section~\ref{subsec:maxdeg5},
we present it here since the basic idea 
is the same one that went into our 2-reductions:
that in some circumstances an optimal assignment of a variable 
is predetermined.
In addition to generalizing from degree 3 to degree~5
(from which the generalization to every degree is obvious),
\cite{clauseLearning} also applies
the idea somewhat differently.

The presentation in \cite{clauseLearning} is specific to 2-Sat.
Reading their result,
it seems unbelievable that it also applies
to \mc as long as the vertex being reduced upon has only light edges
(even if its neighbors have heavy edges),
but in fact the proof carries over unchanged.
For completeness and to make the paper self-contained, 
we present the generalized result.

\newcommand{\np}{N_2^+}
\newcommand{\nn}{N_2^-}
\newcommand{\coll}{C}

\begin{lemma}[clause learning] \label{clause}
In a \mc instance $F$, let $u$ be a variable of degree~5,
with light edges only,
and neighbors $v_1,\ldots,v_5$.
Then there exist ``preferred'' colors $\coll_u$ for $u$
and $\coll_i$ for each neighbor~$v_i$
such that a valid splitting of $F$ is into three instances:
$F_1$ with $\phi(u)=\coll_u$;
$F_2$ with $\phi(u) \neq \coll_u$, $\phi(v_1) = \coll_1$;
and $F_3$ with $\phi(u) \neq \coll_u$,
$\phi(v_1) \neq \coll_1$, 
and $\phi(v_i) = \coll_i$ $(\forall i \in \set{2,3,4,5})$.
\end{lemma}

\begin{proof}
For any coloring $\phi: V \to \bool$,
let $\phi^0$ and $\phi^1$ assign colors 0 and 1 respectively to~$u$,
but assign the same colors as $\phi$ to every other vertex.
That is,
$\phi^i(u)=i$, and $(\forall x\neq u)$ $\phi^i(x)=\phi(x)$.

What we will prove is that for any assignment $\phi$ in which at 
least two neighbors do not 
receive their preferred colors, 
$s(\phi^{\coll_u}) \geq s(\phi)$:
the assignment in which $u$ receives its preferred color 
has score at least as large as that in which it receives the other color,
and thus we may exclude the latter possibility in our search. 
(This may exclude some optimal solutions, but it is also sure to retain 
an optimal solution; thus this trick will not work for counting,
but does work for optimization.)
That is, if $u$ and one neighbor (specifically, $v_1$) do not receive their preferred color, 
then we may assume that every other neighbor receives 
its preferred color.

It suffices to show the existence of colors $\coll_u$
and $\coll_i$, $i \in {1,\ldots,5}$, 
such that for any $\phi$ with $\phi(i) \neq \coll_i$
for two values of $i \in \set{1,\ldots,5}$,
we have $s(\phi^{\coll_u}) \geq s(\phi)$.

\newcommand{\su}{s_u}
\renewcommand{\bias}{b}

Leave the immediate context behind for a moment, 
and consider any \mc instance $F$
in which some variable $u$ 
has only light edges, and in them appears
$\np$ times positively and $\nn$ times negatively.
(As in reduction \ref{2reduce} and Lemma~\ref{super2gen},
we are using the fact that conjunction and disjunction 
are elementwise monotone.)
If $\phi(u)=0$,
the total score $s^0$ from terms involving $u$ satisfies
\begin{align*}
  \su(0) + \nn  & \leq s^0 \leq  \su(0) + \nn + \np ,
 \\ \intertext{%
and if $\phi(u)=1$ the corresponding score $s^1$ satisfies}
  \su(1) + \np  & \leq s^1 \leq  \su(1) + \np + \nn .
\end{align*}
 From the second inequality in the first line
and the first inequality in the second line,
if $\su(1)-\su(0) \geq \nn$ then $s^1 \geq s^0$,
and for any coloring $\phi$, $s(\phi^1) \geq s(\phi^0)$.
Symmetrically, if $\su(0)-\su(1) \geq \np$ then 
$\phi^0$ always dominates $\phi^1$.
Defining the bias 
\begin{align*}
\bias := \su(1)-\su(0) ,
\end{align*}
we may thus infer an optimal color for $u$
if $\bias-\nn \geq 0$ 
or $-\bias-\np \geq 0$.

If $u$ has degree 5, $(\bias-\nn)+(-\bias-\np) = -\nn-\np = -5$,
and thus one of these two parenthesized quantities must be at least $-2.5$,
and by integrality at least~$-2$.
Given the symmetry, \wlg suppose that $\bias-\nn \geq -2$.
The preferred color for $u$ will be $\coll_u=1$.

A small table shows that for any conjunctive or disjunctive clause
involving $u$ or $\bar u$ and some other variable $v_i$
(which \wlg we assume appears positively),
there exists a color $\coll_i$ for $v_i$ (according to the case) such that assigning $v_i$ this color
increases $\bias-\nn$ by~1
(either by increasing the bias and leaving $\nn$ unchanged,
or leaving the bias unchanged and decreasing~$\nn$).
\begin{longtable}{||cc|cccc||}
\hline
\hline
 original & set $\phi(v_i)=$ & resulting & change   & change     & change in \\
 clause   & $C_i=$         & clause    & in $\bias$ & in $\nn$ & $\bias-\nn$ \\
\hline
 $(u \OR v_i)$       & 0 & $(u)$ & $+1$ & $ \phantom- 0$ & $+1$ \\
 $(u \AND v_i)$      & 1 & $(u)$ & $+1$ & $ \phantom- 0$ & $+1$ \\
 $(\bar u \OR v_i)$  & 1 & $(1)$ & $\phantom+ 0$ & $-1$ & $+1$ \\
 $(\bar u \AND v_i)$ & 0 & $(0)$ & $\phantom+ 0$ & $-1$ & $+1$ \\
\hline \hline
\end{longtable}

Thus, starting from $\bias-\nn \geq -2$, 
assigning to any two neighbors of $u$ their color $C_i$
results in an instance in which $\bias-\nn \geq 0$,
and thus in which an optimal assignment for $u$ is $\phi(u)=C_u=1$.
This proves the lemma.
\end{proof}

\subsection{A lemma on 1-reductions}

\newcommand{\Sbar}{{\bar S}}
\newcommand{\Gbar}{{\bar G}}
\newcommand{\disjU}{\uplus}

A half-edge reduction or 1-reduction is ``good'' 
if the target vertex has degree at least $3$,
because (as the weights will come out) 
the measure decrease due to 
$a_{d-1}-a_d$ is substantial for $d \geq 3$,
but small (in fact,~0) for $d=1$ and $d=2$.

If for example we start with a simplified instance 
(in which all vertices must have degree at least $3$)
and reduce on a vertex of degree~d, deleting it and its incident half-edges,
each of the $d$ remaining half-edges implies a good degree reduction
on a neighboring vertex.
However, if we deleted several vertices, this might not be the case:
if two deleted vertices had a common neighbor of degree~3,
its degree would be reduced from 3 to 2 
by one half-edge reduction (good),
but then from 2 to 1 by the other (not good).

The following lemma allows us to argue that a certain number 
of good half-edge reductions occur.
The lemma played a helpful role in our thinking about the case analysis,
but in the presentation here we invoke it rarely:
the cases dealt with are relatively simple, 
and explicit arguments are about as easy as applying the lemma.

Note that for any half edge incident on a vertex~$v$,
we can substitute a full edge between $v$ and 
a newly introduced vertex~$v'$:
after performing a half-edge reduction on $v$ in the first case
or a 1-reduction in the second, the same instance results.
(Also, the measure increase of $a_1$ 
when we add the degree-1 vertex and half-edge 
is canceled by the extra decrease for performing a 1-reduction
rather than a half-edge reduction.)
For clarity of expression, 
the lemma is thus stated in terms of graphs and 1-reductions,
avoiding the awkward half edges.

\begin{lemma} \label{1lemma}
Let $G$ be a graph with $k$ degree-1 vertices, $X=\set{x_1,\ldots,x_k}$.
It is possible to perform a series of 1-reductions in $G$
where each vertex $x_i$ in $X$ is either 
matched one-to-one with a good 1-reduction
(a 1-reduction on a vertex of degree 3 or more),
or belongs to a component of~$G$ containing
at least one other vertex of~$X$,
where the total order of all such components
is at most~$2k$ plus the number of degree-2 vertices.
\end{lemma}

In particular, if $G$ is a connected graph then there are
$k$ good 1-reductions.
By analogy with the well-definedness of the $2$-core of a graph, 
any series of 1-reductions should be equivalent, 
but the weaker statement in the lemma
suffices for our purposes.

\begin{proof}
The intuition is that each series of reductions originating
at some $x_i \in X$, 
after propagating through a series of vertices of degree~2, 
terminates either at a vertex of degree 3 or more 
(reducing its degree), 
establishing a matching between $x$ and a good reduction,
or at another vertex $x_j \in X$, 
in which case the path from $x_i$ to $x_j$ 
(or some more complex structure) is a component.

Starting with~$i=1$, let us 1-reduce from $x_i$ 
as long as possible before moving on to~$x_{i+1}$.
That is, if we 1-reduce into a vertex of degree~2 
we perform a new 1-reduction from that vertex,
terminating when we reach a vertex of degree 1
or degree 3 or more.
Rather than deleting an edge with a 1-reduction, 
imagine that the edges are originally black,
and each reduced edge is replaced by a red one
(which of course is not available for further 1-reductions).

We assert that just before we start processing any~$x_i$,
the red-edged graph has components consisting of vertices
\emph{all} of whose edges are red
(in which case this is also a component in $G$ itself),
and components where all vertices but one \emph{component owner} are all-red,
and the component owner has at least 1 red edge and at least 2 black edges.
We prove this by induction on~$i$, with $i=1$ being trivial.

Given that it is true before~$x_i$, 
we claim that: 
(1)~as we reduce starting with~$x_i$,
the reduction sequence is uniquely determined;
(2)~in the red-edged component including~$x_i$,
all vertices are all-red except for a single \emph{active} one; and
(3)~the sequence on $x_i$ ends when we reduce a vertex
that had at least 3 black edges (matching $x_i$ with this good reduction),
or a vertex $x_j \in X$, $j>i$
(in which case we will show that the red component 
including $x_i$ and $x_j$ is also a component of $G$ itself).

We prove these claims by induction on the step number, 
the base case again being trivial ($x_i$ itself is active).
If we reduce into a vertex $v$ with two black edges 
(we will say it has \emph{black degree~2}), 
the next reduction takes us out its other black edge, 
leaving both red.
If $v$ was of degree~2 it is added to $x_i$'s red component;
if not, it must have been a component owner
(these are the only mixed-color vertices),
and we unite the vertex and its component with $x_i$'s component.
If we reduce into a vertex $v$ with at least 3 black edges,
we match $x_i$ with the good reduction on~$v$,
and $v_i$ owns $x_i$'s red component.
The only remaining possibility is that we reduce into a vertex
with 1 black edge, which 
can only be a degree-1 vertex $x_j$ (with $j>i$),
as there are no mixed-color vertices with 1 black edge. 
In this case we add $x_j$ to $x_i$'s component,
and terminate the sequence of reductions for $x_i$ without a good reduction.
However the red component on $x_i$ now has no black edges on any of 
its vertices, and is thus a component in the original black graph~$G$.

Starting with the $k$ vertices $x_i$ as initial red components, 
as we generate the component for $x_i$, 
the union of all components is expanded 
as we pass through (and use up) a (true) degree-2 vertex,
left unchanged if we pass through a vertex of higher degree
with black degree~2,
expanded as we enter a terminal all-black degree-3 vertex,
and left unchanged if we terminate at another vertex~$x_j$. 
Then,
recalling that $k$ is the number of degree-1 vertices in~$X$
and letting $k_2$ be the number of degree-2 vertices,
the total number of vertices in the union of all components is
at most the number of seeds~($k$),
plus the number of pass-throughs (at most $k_2$), 
plus the number of good terminals (at most~$k$).
In particular, we can partition $X$ into the set of vertices 
with good terminals in~$G$,
and the rest;
the rest lie in components of $G$ 
where the total size of these components is $\leq 2k+k_2$.
\end{proof}

\section{Splitting reductions and preference order}

Recall from \alga that if we have a nonempty simplified instance~$F$, 
we will apply a splitting reduction to produce smaller instances
$F_1,\ldots,F_k$,
simplify each of them,
and argue that $\sum_{i=1}^k \twopow{\mu(F_i)-\mu(F)} \leq 1$
(inequality~\eqref{mux}).

We apply splitting reductions in a prescribed order of preference,
starting with division into components.

\myitem{Split large components} \label{largecomponent}
If the constraint graph $G$ of $F$ has components $G_1$ and~$G_2$
with at least $C$ vertices each
($C$ is the same constant as in 
the simplification rule~\eqref{smallcomponent}),
decompose $F$ into the corresponding instances $F_1$ and~$F_2$.
The decomposition is the obvious one: 
monadic score functions $s_x$ of $F$ are apportioned to $F_1$ or $F_2$
according to whether $x$ is a vertex of $G_1$ or $G_2$, 
similarly for dyadic score functions and edges $xy$,
while we may apportion the niladic score function of $F$ to~$F_1$,
setting that of $F_2$ to~0.

It is clear that this is a valid reduction,
but we must show that \eqref{mux} is satisfied.
Note that $\mua(F_1)+\mua(F_2)=\mua(F)$,
and $\mua(F_i) \geq C a_3$ since $F_i$ has at least $C$ vertices,
all degrees are at least~3, and the $a_i$ are nondecreasing.
Thus $\mua(F_1) \leq \mua(F)- C a_3$.
Also, $\mub(F_1) - \mub(F)$ is constant-bounded.
Assuming that $a_3 > 0$, then for $C$ sufficiently large,
\begin{align*}
 \mu(F_1)-\mu(F) 
   &= \mua(F_1)-\mua(F) + \mub(F_1)-\mub(F) 
   \\ & \leq -C a_3 + \sum_{d=4}^6 (|R_d|+|C_d|)
   \\ & \leq -1 .
\end{align*}
The same is of course true for $F_2$, giving 
$ \twopow{\mu(F_1)-\mu(F)} + \twopow{\mu(F_2)-\mu(F)}
 \leq 2^{-1} + 2^{-1} = 1$
as required.

The non-strict inequality $a_3 \geq 0$ is established by~\eqref{apos},
and if $a_3=0$, a $3$-regular (cubic) instance would have measure~0,
implying that we could solve it in polynomial time,
which we do not know how to do.
Thus 
let us assume for a moment that
\begin{align}
a_3 &> 0 . \label{a3posstrict}
\end{align}
This strict inequality (in fact $a_3 \geq 1/7$) will be implied by
the constraints for splitting rules for cubic instances, constraint \eqref{3cut} for example.
\caseend

If $F$'s constraint graph is connected
the splitting we apply depends on the degree of~$F$,
that is, the degree of its highest-degree vertex.
Although high-degree cases thus take precedence,
it is easier to discuss the low-degree cases first.
Sections \ref{subsec:cubic}, 
\ref{subsec:maxdeg4}, \ref{subsec:maxdeg5}, and~\ref{subsec:maxdeg6}
detail the splittings for (respectively) instances 
of degree 3, 4, 5, and~6.
For a given degree,
we present the reductions in order of priority.

\section{Cubic instances}
\label{subsec:cubic}

Many formulas are not subject to any of the simplification rules above
nor to large-component splitting.
In this section we introduce further reductions so that for
any formula $F$ of maximum degree at most~3
(which is to say, whose constraint graph has maximum degree at most~3),
some reduction can be applied.

If $F$ has any vertex of degree strictly less than~3, 
we may apply the 0-, 1-, or 2-reductions above.
Henceforth, then, we assume that $F$ is 3-regular (cubic).

The new reductions will generally be ``atomic''
in the sense that we will carry each through to its stated completion,
not checking at any intermediate stage whether 
an earlier simplification or reduction rule can be applied.

We define 
\begin{align}
 h_3 & := a_3 - a_2  \label{h3def}
\end{align}
to be the decrease of measure resulting from
a half-edge reduction \myref{halfedge} on a vertex of degree~$3$.

\newcommand{\VSX}{V \setminus \set{S \cup X}}

\myitem{3-cut} \label{Case3cut}  
There is a $3$-cut $X=\{x_1,x_2,x_3\}$ 
isolating a set $S$ of vertices, with $4 \leq |S| \leq 10$.
Each cut vertex $x_i$ has at least 1 neighbor in $\VSX$
(otherwise $X$ without this vertex is a smaller cut),
and without loss of generality we may assume that 
either each cut vertex has 2 neighbors in $\VSX$,
or that $|S|=10$.
(If a cut vertex, say $x_1$, 
has just one neighbor $x'_1 \in \VSX$,
then $\set{x'_1, x_2, x_3}$ is also a 3-cut,
isolating the larger set $S \cup \set{x_1}$.
Repeat until $|S|=10$ or each cut vertex has two neighbors in $\VSX$.)

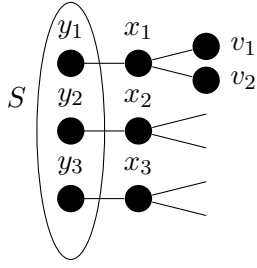
\begin{figure}[tbp]
       \centering
       \begin{tikzpicture}[scale=0.9]
        \tikzstyle{vertex}=[minimum size=1mm,circle,fill=black]
        
        \draw (0,1) ellipse (0.5 and 1.9);
        \draw (-0.8,1.5) node {$S$};
        \draw (1,2) node[vertex,label=above:$x_1$] (x1) {};
        \draw (1,1) node[vertex,label=above:$x_2$] (x2) {};
        \draw (1,0) node[vertex,label=above:$x_3$] (x3) {};
        \draw (0,2) node[vertex,label=above:$y_1$] (y1) {};
        \draw (0,1) node[vertex,label=above:$y_2$] (y2) {};
        \draw (0,0) node[vertex,label=above:$y_3$] (y3) {};
        \draw (2,2.25) node[vertex,label=right:$v_1$] (y1) {};
        \draw (2,1.75) node[vertex,label=right:$v_2$] (y2) {};

        \draw (0,2)--(x1) (0,1)--(x2) (0,0)--(x3)
              (x1)--(y1) (x1)--(y2) (x2)--(2,1.25) (x2)--(2,0.75) (x3)--(2,0.25) (x3)--(2,-0.25) ;
       \end{tikzpicture}
       \caption{Illustration of a 3-cut, reduction~\ref{Case3cut}. 
         \label{figCase11}}
\end{figure}

With reference to Figure~\ref{figCase11},
let $y_1, y_2, y_3 \in S$ be the respective neighbors of
$x_1$, $x_2$, and~$x_3$,
and let $v_1$ and $v_2$ be the other neighbors of~$x_1$.
Note that $y_2 \neq y_3$, or we should instead apply a 2-cut reduction
\myref{2cut}:
cutting on $\set{x_1,y_2}$ 
isolates the set $S \setminus \set{y_2}$,
and $3 \leq |S \setminus \set{y_2}| \leq 9$ satisfies
the conditions of the 2-cut reduction.

We treat this case by splitting on $x_1$,
resulting in new instances $F_1$ and~$F_2$.
In each we apply a 2-cut on $\set{y_2,y_3}$
(not $\set{x_2,x_3}$!),
creating a possibly-heavy edge~$y_2 y_3$.
We then 2-reduce on $y_2$ and $y_3$ in turn 
to create an edge $x_2 x_3$ which is heavy only if
$x_2 y_2$ and $x_3 y_3$ were both heavy.
If $|S| \leq 10$, the resulting instances satisfy
\begin{align*} 
 \mu(F_1), \mu(F_2) 
   \leq \mu(F) - 5 a_3 - 2 h_3 .
\end{align*}
(Recall that for graphs of degree~3, $\mu$ and $\mua$ are identical.)
The term $-5 a_3$ accounts for the deletion of $x_1$ and $S$
(at least 5 vertices) with their incident half-edges.
The term $-2 h_3$ accounts for deletion of the ``other halves''
of the edges from $x_1$ to $\VSX$
and the degree decrease of their incident vertices
(see definition~\eqref{h3def});
we are using the fact that $v_1 \neq v_2$,
and that $X$ is an independent set.
There is no need for a term accounting for the deletion of the 
``other halves'' of the edges on $x_2$ and $x_3$
and the addition of the new edge $x_2 x_3$:
the new $x_2 x_3$ is heavy only if both half-edges were heavy,
so this change in measure is
$-\tfrac12 w(x_2 y_2) -\tfrac12 w(x_3 y_3) + w(x_2 x_3) \leq 0$,
and we are free to ignore it.
(Since it may in fact be~0, there is also no gain to including it.)
Constraint \eqref{mux} of Lemma~\eqref{mainlemma} is thus assured if
\begin{align*}
 \twopow{- 5 a_3 - 2 h_3 }
 + \twopow{- 5 a_3 - 2 h_3 }
 \leq 2^0 = 1 .
\end{align*}
We will henceforth express such constraints by a shorthand, 
simply saying that the case has splitting number at most
\begin{align}
 (5 a_3 + 2 h_3, 
    5 a_3 + 2 h_3 ) . \label{3cut}
\end{align}
We formally define a \emph{splitting number} to be
\begin{align*}
 (\alpha_1, \alpha_2, \dots, \alpha_k) := 2^{- \alpha_1} + 2^{- \alpha_2} + \dots + 2^{- \alpha_k} .
\end{align*}
Note the change of sign: 
in this notation we show the cost \emph{decrease} in each case.

By similar reasoning, if $|S|=10$ the splitting number is at most
\begin{align*}
 (  11 a_3 + h_3 ,
    11 a_3 + h_3 ) .
\end{align*}
By \eqref{a3posstrict} this constraint is bound to hold
``for a sufficiently large value of 10''
(and since $h_3 \leq a_3$, for 10 itself this constraint
is dominated by~\eqref{3cut}),
so we will disregard it.
\caseend

\newcommand{\Nbar}{{\bar N}}

\myitem{Vertex with independent neighbors} \label{Case1p3} 
There is a vertex $u$ such that $N(u)$ is an independent set.
 
\begin{figure}[htbp]
       \centering
       \begin{tikzpicture}[scale=1]
        \tikzstyle{vertex}=[minimum size=1mm,circle,fill=black]
        \draw (0,1) node[vertex,label=left:$u$] (u) {};
        \draw (1,2) node[vertex,label=above:$v_1$] (v1) {};
        \draw (1,1) node[vertex,label=above:$v_2$] (v2) {};
        \draw (1,0) node[vertex,label=above:$v_3$] (v3) {};
        \draw (2,2.5) node[vertex] (x1) {};
        \draw (2,1.75) node[vertex] (x2) {};
        \draw (2,1.25) node[vertex,label=right:$x_3$] (x3) {};
        \draw (2,0.5) node[vertex,label=right:$x_4$] (x4) {};
        \draw (2,-0.25) node[vertex] (x5) {};

        \draw (u)--(v1) (u)--(v2) (u)--(v3)
              (v1)--(x2) (v2)--(x3) (v2)--(x4) (v3)--(x4) (v3)--(x5);
        \draw[very thick, double] (v1)--(x1);
       \end{tikzpicture}
       \caption{Illustration for reduction~\ref{Case1p3},
         on a vertex with independent neighbors.
         \label{figCase1321}
       }
      \end{figure}
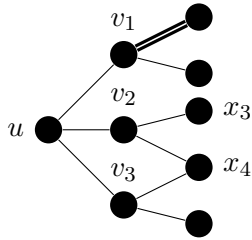

With reference to Figure~\ref{figCase1321},
we reduce on~$u$, fixing $\phi(u)$ to $0$ and $1$ 
to generate new instances $F_0$ and~$F_1$,
each with constraint graph $G[V \setminus \set{u}]$.

Let $N^1 = N(u)$ and $N^2 = N^2(u)$.
Let $q$ be the number of vertices in $N^1$ with a heavy edge to~$N^2$, 
$k_0$ the number of vertices in $N^1$ subject to a super 2-reduction
(deletion) in~$F_0$,
and $k_1$ the number subject to super 2-reduction in~$F_1$.
By Lemma~\ref{super2gen}, 
each $v \in N^1$ falls into at least one of these cases,
so $q+k_0+k_1 \geq 3$.

We will argue that $\mu(F)-\mu(F_i) \geq a_3 + 3h_3 + q (w_h-w_e) + 2 k_i h_3$.
Deletion of $u$ and reduction of the degree of each of its neighbors
immediately reduces the measure by $a_3 + 3 h_3$
(more if any edges incident to $u$ were heavy).
In~$F_i$, first 2-reduce on the $q$ vertices in $N^1$ with heavy edges
(reducing the measure by a further $q (w_h-w_e)$)
and on the $3-q-k_i$ vertices subject to only plain 2-reductions
(not increasing the measure).
Note that each vertex in $N^2$ still has degree~3.

Finally, reduce out the $k_i$ vertices which are set constant
by a super 2-reduction, 
by deleting their incident edges one by one.
No vertex $v$ in $N^2$ has 3 neighbors in $N^1$:
if it did there would
remain only 3 other edges from $N^1$ to~$N^2$,
whence $|N^2| \leq 4$,
$N^2 \setminus v$ would be a cut of size $\leq 3$ isolating 
$N^1 \cup \set{u,v}$, and we would have applied a cut reduction.
Thus, 
deletion of each of the $2k_i$ edges in $N^1 \times N^2$ 
either reduces the degree of a vertex in $N^2$ from 3 to~2
(a good 1-reduction, reducing the measure by~$h_3$),
or creates a vertex of degree~1.

We wish to show that each degree-1 vertex 
in the graph $G'=G[V \setminus (\set u \cup N^1)]$
must also result in a good 1-reduction, giving the $2 k_i h_3$ claimed.
Note that $|N^2|$ must be 4, 5, or~6
(if it were smaller we would have applied a cut reduction instead).
If $|N^2|=6$ then every vertex in $N^2$ 
has degree~2 (in the graph~$G'$)
and there is nothing to prove.
If $|N^2|=5$ then at most one vertex in $N^2$ has degree~1,
and Lemma~\ref{1lemma} implies that it results in a good 1-reduction.
If $|N^2|=4$, every degree-1 vertex in $N^2$ also results in a 
good 1-reduction:
If not, then by Lemma~\ref{1lemma}
a set $X$ of two or more vertices in $N^2$ 
lies in a small component of~$G'$,
in which case $N^2 \setminus X$ is a cut of size 2 or less in
the original constraint graph~$G$,
isolating $\set{u} \cup N^1 \cup X$, 
and we would have applied a cut reduction instead.

Thus, $\mu(F)-\mu(F_i) \geq a_3+3 h_3+q (w_h-w_e) +2 k_i h_3$.
By convexity, if two splitting numbers have equal total,
the more unbalanced one is the more constraining;
in this case that means the worst cases come if $k_0=0$
and $k_1=3-q$ (or vice-versa).
Thus, the worst-case splitting numbers are
\begin{align}\label{indep3}
 (\forall q \in \set{0,1,2,3}) \quad
     (a_3+3 h_3 + q (w_h-w_e),
      a_3+3 h_3 + q (w_h-w_e) + 2 (3-q) h_3) .
\end{align}
\caseend

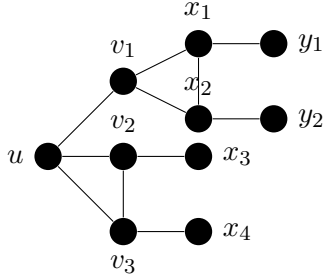
\begin{figure}[tbp]
       \centering
       \begin{tikzpicture}[scale=1]
        \tikzstyle{vertex}=[minimum size=1mm,circle,fill=black]
        \draw (0,1) node[vertex,label=left:$u$] (u) {};
        \draw (1,2) node[vertex,label=above:$v_1$] (v1) {};
        \draw (1,1) node[vertex,label=above:$v_2$] (v2) {};
        \draw (1,0) node[vertex,label=below:$v_3$] (v3) {};
        \draw (2,2.5) node[vertex,label=above:$x_1$] (x1) {};
        \draw (2,1.5) node[vertex,label=above:$x_2$] (x2) {};
        \draw (2,1) node[vertex,label=right:$x_3$] (x3) {};
        \draw (2,0) node[vertex,label=right:$x_4$] (x4) {};
        \draw (3,2.5) node[vertex,label=right:$y_1$] (y1) {};
        \draw (3,1.5) node[vertex,label=right:$y_2$] (y2) {};

        \draw (u)--(v1) (u)--(v2) (u)--(v3)
              (v2)--(v3)
              (v1)--(x1) (v1)--(x2) (v2)--(x3) (v3)--(x4)
              (x1)--(x2)
              (x1)--(y1) (x2)--(y2);
       \end{tikzpicture}
       \caption{Illustration of reduction on a vertex with 
         one edge in its neighborhood, Case~\ref{Case1p4}.
         \label{figCase14}
       }
      \end{figure}

\myitem{One edge in $G[N(u)]$} \label{Case1p4} 
Given that we are in this case rather than Case~\ref{Case1p3},
no vertex of $N(u)$ has an independent set as neighborhood.
Let $N(u)=\{v_1,v_2,v_3\}$ and suppose \wlg that $v_2 v_3 \in E$. 
Let $N(v_1)=\{u,x_1,x_2\}$. 
Then, $x_1 x_2 \in E$. 
To avoid a 3-cut (Case~\ref{Case3cut}), $|N^2(\{u,v_1\})|=4$
(the 4 rightmost vertices depicted in Figure~\ref{figCase14}
are truly distinct).
      
After splitting on $u$, in each of the two instances $F_0$ and~$F_1$,
first 2-reduce on $v_1$, then on $x_1$,
then continue with 2-reductions (the base case),
or super 2-reductions (if possible), on $v_2$ and~$v_3$.
In the base case this
results in the deletion of all 5 of these vertices 
with their incident edges and the decrease of the degree
of $x_2$ to $2$, for a measure decrease of~$5 a_3 + h_3$
(vertex $x_2$ will be 2-reduced, which does not increase the
measure; see \ref{2reduce}).

If $v_2 v_3$ or $v_2 x_3$ is heavy, then there is an extra 
measure decrease of $w_h-w_e$ beyond that of the base case,
for a splitting number of at most
\begin{align}\label{edge31}
 ( 5 a_3 + h_3 + w_h - w_e , 5 a_3 + h_3 + w_h - w_e ) .
\end{align}

Otherwise, $v_2 v_3$ and $v_2 x_3$ are both light, and
we may super 2-reduce on $v_2$ in either $F_0$ or~$F_1$
(\wlg say $F_1$).
This reduces the degree of $x_3$ from 3 to~2,
and that of $v_3$ from 2 to~1,
setting up a 1-reduction on $v_3$ 
that reduces the degree of $x_4$ from 3 to~2.
This gives a splitting number of at most
\begin{align}\label{edge32}
 ( 5 a_3 + h_3 , 5 a_3 + 3 h_3 ) .
\end{align}
\caseend

There are no further cases for cubic graphs. 
If for a vertex $u$ there are 3 edges in $G[N(u)]$ then 
$N[u]$ is an isolated component 
(a complete graph~$K_4$) and we apply component-splitting.
If there are 2 edges in $G[N(u)]$, 
then some $v \in N(u)$ 
(either of the vertices having a neighbor outside $\set{u} \cup N(u)$)
has just 1 edge in $G[N(v)]$ and we are back to Case~\ref{Case1p4}.

\subsection{Cubic results}

For results on cubic and other instances, we 
refer to Theorem~\ref{thm:runtimes},
Table~\ref{tab:runtimes},
and the discussion in Section~\ref{tuning}.

\subsection{Remark on heavy edges}
If the original cubic instance is a pure 2-Sat formula,
with no heavy edges, then 
(as we show momentarily) any heavy edges introduced by 
the procedure we have described can immediately be removed.
Thus the ``hybrid formula'' concept gives no gain for cubic 2-Sat formulas,
but expands the scope to cubic \mc, sacrifices nothing,
and is useful for analyzing non-cubic instances.
We now show how heavy edges introduced into a cubic 2-Sat formula
immediately disappear again.

In a graph with only light edges,
the only two rules that create heavy edges are 2-reductions 
and 2-cuts (and other reductions that apply these).
A 2-reduction on $v$ introduces a heavy edge only if $v$'s neighbors
$x_1$ and $x_2$ were already joined by an edge.
In that case, though, $x_1$ and $x_2$ have their degrees
reduced to 2 (at most).
If the remaining neighbors $y_1$ of $x_1$ and $y_2$ of $x_2$ 
are distinct, then 2-reducing on $x_1$ gives a light edge $x_2 y_1$:
the heavy edge $x_1 x_2$ is gone.
Otherwise, $y_1=y_2$, and 2-reduction on $x_1$ followed by 1-reduction
on $x_2$ deletes $x_1$ and $x_2$ and reduces the degree of $y_2$ to~1,
again leaving no heavy edge.

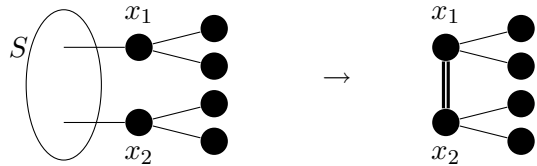
\begin{figure}[htbp]
       \centering
       \begin{tikzpicture}[scale=1, baseline=(ctre)]
        \tikzstyle{vertex}=[minimum size=1mm,circle,fill=black]
        
        \node (ctre) at (0,1) {};
        \draw (ctre) ellipse (0.5 and 1);
        \draw (-0.6,1.5) node {$S$};
        \draw (1,1.5) node[vertex,label=above:$x_1$] (x1) {};
        \draw (1,0.5) node[vertex,label=below:$x_2$] (x2) {};
        \draw (2,1.75) node[vertex] (y1) {};
        \draw (2,1.25) node[vertex] (y2) {};
        \draw (2,0.75) node[vertex] (y3) {};
        \draw (2,0.25) node[vertex] (y4) {};

        \draw (0,1.5)--(x1) (0,0.5)--(x2)
              (x1)--(y1) (x1)--(y2) (x2)--(y3) (x2)--(y4);
       \end{tikzpicture}
       \hspace{1cm}
       $\to$
       \begin{tikzpicture}[scale=1, baseline=(ctre)]
        \tikzstyle{vertex}=[minimum size=1mm,circle,fill=black]
        
        \node (ctre) at (0,1) {};
        \draw (1,1.5) node[vertex,label=above:$x_1$] (x1) {};
        \draw (1,0.5) node[vertex,label=below:$x_2$] (x2) {};
        \draw (2,1.75) node[vertex] (y1) {};
        \draw (2,1.25) node[vertex] (y2) {};
        \draw (2,0.75) node[vertex] (y3) {};
        \draw (2,0.25) node[vertex] (y4) {};

        \draw (x1)--(y1) (x1)--(y2) (x2)--(y3) (x2)--(y4);
        \draw[very thick, double] (x1)--(x2);        
       \end{tikzpicture}
       
       \caption{$2$-cut rule creates a heavy edge.
        \label{bad2cut}}
      \end{figure}

For a 2-cut on $x_1$ and $x_2$ isolating a set~$S$,
if there was an edge $x_1 x_2$ then the cut reduction reduces the
degrees of both $x_1$ and $x_2$ to~2, and, just as above, 
we may 2-reduce on $x_1$ to eliminate the heavy edge.
If $x_1$ and $x_2$ are nonadjacent and $x_1$ has 
just 1 neighbor outside~$S$, 
then again a follow-up 2-reduction on $x_1$ 
eliminates the heavy edge~$x_1 x_2$.
Dismissing the symmetric case for~$x_2$, 
all that remains is the case when $x_1$ and $x_2$ are nonadjacent
and each has 2 neighbors outside~$S$,
and thus just 1 neighbor in~$S$; see Figure~\ref{bad2cut}.

The $S$-neighbors $x'_1$ of $x_1$ and $x'_2$ of $x_2$ must be distinct,
or else we would have applied a 1-cut reduction on~$x'_1$.
(This presumes that $|S \setminus \set{x'_1}| \geq 2$,
but if it is 0 or~1,
we would have 2-reduced on $x'_1$
or 1-reduced on its $S$-neighbor ---
either of which is really a special case of a 1-cut reduction.)

\begin{figure}[htbp]
       \centering
       \begin{tikzpicture}[scale=1, baseline=(ctre)]
        \tikzstyle{vertex}=[minimum size=1mm,circle,fill=black]

        \node (ctre) at (0,1) {};
        \draw (ctre) ellipse (0.5 and 1);
        \draw (-0.6,1.5) node {$S$};
        \draw (1,1.5) node[vertex,label=above:$x_1$] (x1) {};
        \draw (1,0.5) node[vertex,label=below:$x_2$] (x2) {};
        \draw (0.25,1.5) node[vertex,label=left:$x_1'$] (x1') {};
        \draw (0.25,0.5) node[vertex,label=left:$x_2'$] (x2') {};
        \draw (2,1.75) node[vertex] (y1) {};
        \draw (2,1.25) node[vertex] (y2) {};
        \draw (2,0.75) node[vertex] (y3) {};
        \draw (2,0.25) node[vertex] (y4) {};

        \draw (x1')--(x1) (x2')--(x2)
              (x1)--(y1) (x1)--(y2) (x2)--(y3) (x2)--(y4);
       \end{tikzpicture}
       \hspace{1cm}
       $\to$
       \hspace{0.5cm}
       \begin{tikzpicture}[scale=1, baseline=(ctre)]
        \tikzstyle{vertex}=[minimum size=1mm,circle,fill=black]
        
        \node (ctre) at (0,1) {};
        \draw[transparent] (ctre) ellipse (0.5 and 1);
        \draw (1,1.5) node[vertex,label=above:$x_1$] (x1) {};
        \draw (1,0.5) node[vertex,label=below:$x_2$] (x2) {};
        \draw (0.25,1.5) node[vertex,label=left:$x_1'$] (x1') {};
        \draw (0.25,0.5) node[vertex,label=left:$x_2'$] (x2') {};
        \draw (2,1.75) node[vertex] (y1) {};
        \draw (2,1.25) node[vertex] (y2) {};
        \draw (2,0.75) node[vertex] (y3) {};
        \draw (2,0.25) node[vertex] (y4) {};

        \draw (x1')--(x1) (x2')--(x2)
              (x1)--(y1) (x1)--(y2) (x2)--(y3) (x2)--(y4);
        \draw[very thick, double] (x1')--(x2');        
       \end{tikzpicture}
       \hspace{1cm}
       $\to$
       \begin{tikzpicture}[scale=1, baseline=(ctre)]
        \tikzstyle{vertex}=[minimum size=1mm,circle,fill=black]
        
        \node (ctre) at (0,1) {};
        \draw (1,1.5) node[vertex,label=above:$x_1$] (x1) {};
        \draw (1,0.5) node[vertex,label=below:$x_2$] (x2) {};
        \draw (2,1.75) node[vertex] (y1) {};
        \draw (2,1.25) node[vertex] (y2) {};
        \draw (2,0.75) node[vertex] (y3) {};
        \draw (2,0.25) node[vertex] (y4) {};

        \draw (x1)--(x2)
              (x1)--(y1) (x1)--(y2) (x2)--(y3) (x2)--(y4);     
       \end{tikzpicture}
       
       \caption{$2$-cut rule avoids creating a heavy edge.
        \label{good2cut}}
      \end{figure}
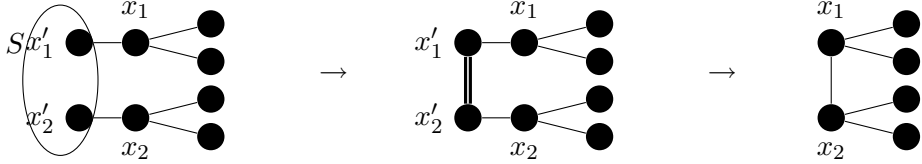

Given that $x'_1 \neq x'_2$,
apply a 2-cut reduction not on $x_1$ and $x_2$
but instead on $x'_1$ and $x'_2$.
Following this with 2-reduction on $x'_1$ and $x'_2$
eliminates the heavy edge $x'_1 x'_2$,
giving a light edge $x_1 x_2$ instead;
see Figure~\ref{good2cut}.

\subsection{Solving the programs} \label{solver}
Every weight constraint we introduce is of the form 
$\sum_i \twopow{L_i} \leq 1$,
where the sum is finite and each $L_i$ is some linear combination of weights. 
(Some constraints are simply of the form $L \leq 0$,
but this can also be written as $\twopow L \leq 1$.)
This standard form 
(along with the objective of minimizing~$w_e$)
can be provided, through an interface such as AMPL,
to a variety of mathematical-programming solvers:
we used both IPOPT 
(part of the free, open-source code repository at \texttt{coin-or.org})
and MINOS (a commercial solver).

Furthermore, it is easily verified that the feasible region is convex.
(Convexity of $2^x$ means that for any $p,q\geq 0$, with $p+q=1$, term by term, 
$\twopow{pL + qL'} \leq p \twopow{L}+ q \twopow{L'}$,
and thus a mixture of feasible solutions is feasible.)
This in turn makes it relatively easy for a solver to 
return a provably optimal solution:
convex programs are much easier to solve than general ones
or even the quasi-convex programs like Eppstein's~\cite{Eppstein}.

IPOPT solves the nonlinear program for our general algorithm,
to optimality, in a second or two on a typical laptop computer.

To insure that our solutions are truly feasible, 
in the presence of finite numerical accuracy,
we replace the ``1'' in the right-hand side of each constraint 
with $1-\epsilon$, fixing $\epsilon=10^{-6}$;
this allows some margin for error.
The values we show for the key parameters $w_e$ and $w_h$
are rounded up (pessimistically) 
 from the higher-precision values returned by the solver,
with the other parameter values rounded fairly.
Ideally we would also verify,
in an unlimited-accuracy tool such as Mathematica,
that our rounded values satisfy the original ``$\leq 1$''
constraints, but we have not performed that final check.

\section{Instances of degree 4}
\label{subsec:maxdeg4}

We first introduce one more bit of notation,
generalizing our earlier definition of $h_3$ \eqref{h3def}.
For any $d \geq 3$, we define 
\begin{align}
 h_d &:= \min_{3 \le i \le d} \{ 
         a_i - a_{i-1} 
        \} . 
  \label{halfred}
\end{align}
This is the minimum possible decrease of measure resulting from 
a half-edge reduction \myref{halfedge} on a vertex of degree~$i$
with $3 \leq i \leq d$.
We will find that such deletions always occur with the same sign
in our nonlinear program 
--- the larger $h_d$, the weaker each constraint is ---
and therefore the above definition can be expressed in
our mathematical program 
by simple inequalities 
\begin{align}
 (\forall 3 \le i \le d) \quad 
 h_d & \leq   
 a_i - a_{i-1} .
\end{align}

We now consider a formula $F$ of (maximum) degree~$4$.
The algorithm choses a vertex $u$ of degree $4$ with 
--- if possible --- at least
one neighbor of degree~$3$. 
The algorithm sets $u$ to $0$ and~$1$, 
simplifies each instance as much as possible 
(see Section~\ref{simplification}),
and recursively solves the resulting instances $F_0$ and~$F_1$.

The instances $F_0$ and $F_1$ are either $4$-regular,
of degree at most~$3$, or nonregular.
By the arguments presented in Section~\ref{measuresform},
the case where the degree of the graph decreases can be
safely ignored 
(the measure decrease $C_4 - C_3$ can be made as large as necessary).

\myitem{4-regular} \label{CaseReg4}
If $F$ is 4-regular, 
first consider the case in which
$F_0$ and $F_1$ are $4$-regular. 
Since splitting on $u$ decreases the degree of each vertex in $N(u)$, 
and none of our reduction rules increases the degree of a vertex,
every vertex in $N(u)$ must have been removed from $F_0$ and $F_1$ 
by simplification rules.%
\footnote{%
There is an important subtlety here:
the reduced-degree vertices are eliminated,
not merely split off into other components
such that $F_i$ has a 4-regular component and a component of degree~3
(although such an example shares with 4-regularity
the salient property that no degree-4 vertex has a degree-3 neighbor).
By definition, the ``4-regular case'' we are considering at this point
does not include such an $F_i$,
but it is worth thinking about what happens to an $F_i$ 
which is not regular but has regular components.
No component of $F_i$ is small
(simplification \ref{smallcomponent} has been applied),
so in the recursive solution of~$F_i$, 
\alga immediately applies large-component splitting \myref{largecomponent}.
This reduces $F_i$ to two connected instances, and
is guaranteed to satisfy constraint~\eqref{mux}
(the penalty for one instance's being 4-regular
is more than offset by its being much smaller than~$F_i$).
Our machinery takes care of all of this automatically, 
but the example illustrates why some of the machinery is needed.%
}
This gives a splitting number of at most
\begin{align}\label{splittingReg41}
 \left( 5 a_4, 5 a_4 \right) .
\end{align}

If neither $F_0$ nor $F_1$ is 4-regular, 
then $u$ is removed ($a_4$),
the degree of its neighbors decreases ($4 h_4$), 
and we obtain an additional gain because
$F_0$ and $F_1$ are not regular ($R_4$). 
Thus, the splitting number is at most
\begin{align}\label{splittingReg42}
 \left( a_4 + 4 h_4 + R_4, a_4 + 4 h_4 + R_4 \right) .
\end{align}

If exactly one of $F_0$ and $F_1$ is 4-regular, 
we obtain a splitting number of
$ \left( 5 a_4 , a_4 + 4 h_4 + R_4 \right) $.
This constraint is weaker (no stronger) than 
\eqref{splittingReg41}
if $5 a_4 \leq a_4 + 4 h_4 + R_4$,
and weaker than 
\eqref{splittingReg42}
if $5 a_4 > a_4 + 4 h_4 + R_4$,
so we may dispense with it.
%

\myitem{4-nonregular} \label{CaseNonReg4}
If $F$ is not 4-regular, 
we may assume that $u$ has at least one neighbor of degree~$3$. 
Let us denote by $p_i$ the number degree-$i$ neighbors of~$u$. 
Thus, $1 \le p_3 \le 4$, and $p_3 + p_4 = 4$.
Further, let us partition the set $P_3$ of degree-3 neighbors 
into those incident only to light edges, $P_3'$,
and those incident to at least one heavy edge, $P_3''$.
Define $p_3' = |P_3'|$ and $p_3''=|P_3''|$
(so $p_3'+p_3''=p_3$).

For each $F_i$ ($F_0$ and $F_1$),
splitting on $u$ removes $u$ 
(for a measure decrease of $a_4$, compared with~$F$).
If $F_i$ is not 4-regular,
the degrees of the neighbors of $u$ all decrease ($\sum_{i=3}^4 p_i h_i$).
If $F_i$ is regular ($- R_4$), 
all neighbors of $u$ must have been eliminated as well
($\sum_{i=3}^4 p_i a_i$).

We now argue about additional gains based on the values of $p_3'$ and $p_3''$,
starting with the heavy edges incident on vertices in $P_3''$.
Identify one heavy edge on each such vertex.
If such an edge is between two vertices in $P''_3$ associate it with either 
one of them; otherwise associate it with its unique endpoint in $P''_3$.
This gives a set of at least $\ceil{p''_3/2}$ vertices in $P''_3$
each with a distinct associated heavy edge,
which we may think of as oriented out of that vertex. 
If such an edge incident on $v \in P''_3$
is also incident on $u$ then it is deleted along with~$u$,
for an additional measure reduction of $w_h-w_e$ we credit to~$v$.
This leaves a set of ``out'' edges that may form paths or cycles. 
After deletion of $u$ all the vertices involved have degree~2,
so any cycle is deleted as an isolated component, 
for a measure reduction of $w_h-w_e$ per vertex.
Super 2-reducing on a vertex $v$ deletes its outgoing edge,
which we credit to~$v$,
and possibly also an incoming heavy edge associated with a different
$v' \in P''_3$, which we credit to~$v'$. 
Finally, if $v$ is 2-reduced we consider its outgoing edge 
(not its other incident edge) to be contracted out along with~$v$,
crediting this to~$v$
(and correctly resulting in a light edge if the other edge incident on $v$
was light, or a heavy one if it was heavy).
This means that if the other edge incident to $v$ was a heavy edge
out of a different $v' \in P''_3$,
then $v'$ still has an associated outgoing heavy edge. 
In short, each of the $\ceil{p''_3/2}$ vertices gets credited with the
loss of a heavy edge, for an additional measure reduction of
at least $\ceil{p''_3/2} (w_h-w_e)$.

We say that we have a \emph{good degree reduction}
if the degree of a vertex of degree 3 or more decreases by~1:
for graphs of degree 4 this decreases the measure by at least~$h_4$.
This measure decrease comes in addition to what we have accounted 
for so far, 
unless $F_i$ is regular and the degree reduction is on a vertex in $N(u)$
(since we have accounted for the deletion of those vertices,
counting their degree reductions as well would be double counting).
We will show that a certain number of additional-scoring degree reductions
occur altogether, in $F_0$ and $F_1$ combined, as a function of~$p_3'$.

If $p_3' = 1$, super 2-reduction on the sole vertex in $P_3'$
is possible in at least one of $F_0$ or $F_1$
--- without loss of generality say just $F_0$ --- and 
reduces the degrees of at least two neighbors. 
If $F_0$ is nonregular this gives a gain of $2 h_4$,
while if $F_0$ is regular there may be no gain.

\begin{figure}[htbp]
       \centering
       \begin{tikzpicture}[scale=0.7]
        \tikzstyle{vertex}=[minimum size=1mm,circle,fill=black]
        \draw (0,0) node[vertex,label=left:$u$] (u) {};
        \draw (2,1.5) node[vertex,label=above:$v_1$] (v1) {};
        \draw (2,0.5) node[vertex,label=right:$v_2$] (v2) {};
        \draw (2,-0.5) node[vertex,label=below right:$v_3$] (v3) {};
        \draw (2,-1.5) node[vertex,label=below:$v_4$] (v4) {};
        \draw (4,1) node[vertex,label=right:$x_1$] (x1) {};
        \draw (4,0) node[vertex,label=right:$x_2$] (x2) {};
        \draw (4,-1) node[vertex,label=right:$x_3$] (x3) {};

        \draw (u)--(v1) (u)--(v2) (u)--(v3) (u)--(v4)
	      (v1)--(v2) (v2)--(v3) 
              (v1)--(x1) (v3)--(x2) (v4)--(x3);
        \draw[very thick, double] (v3)--(v4);
       \end{tikzpicture}
       \caption{The case $p_3'=2$ may lead to just one good degree reduction
       outside $N[u]$.
	If both super 2-reductions on $v_1$ and $v_2$ occur in the same branch (say $F_1$),
	the degree of~$x_1$ is reduced.
	The degrees of $v_3$ and $v_4$ become~2, so their edges are
	contracted eventually creating an edge $x_2 x_3$, which 
	does not change the degree of $x_2$ or~$x_3$.
	The heavy edge $v_3 v_4$ gives a bonus measure reduction 
	of $w_h-w_e$ previously accounted for.
         \label{OneGoodRed}
       }
\end{figure}
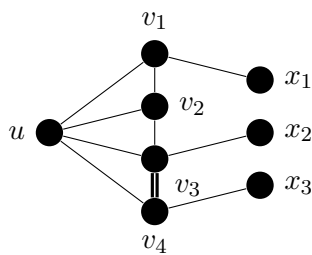

If $p_3' = 2$, then again if either vertex is super 2-reduced 
in a nonregular branch there is a gain of at least $2 h_4$.
Otherwise, each vertex is super 2-reduced in a regular branch
(both in one branch, or in two different branches, as the case may be). 
At least one of the vertices has at least one neighbor in $N^2 := N^2(G)$,
or else $P_3 \setminus P'_3$ would be 2-cut.
In whichever $F_i$ the degree of the neighbor is reduced,
since $F_i$ is regular the neighbor must eventually be deleted,
for a gain of at least $a_3$. 
So there is either a gain of $2 h_4$ in a nonregular branch
or a gain of $a_3$ in a regular branch.
(We cannot hope to replace $a_3$ with $2 a_3$:
Figure~\ref{OneGoodRed} shows an example where indeed only one
good degree reduction occurs outside $N[u]$.)

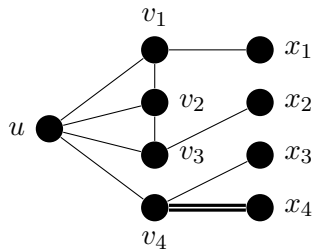
\begin{figure}[htbp]
       \centering
       \begin{tikzpicture}[scale=0.7]
        \tikzstyle{vertex}=[minimum size=1mm,circle,fill=black]
        \draw (0,0) node[vertex,label=left:$u$] (u) {};
        \draw (2,1.5) node[vertex,label=above:$v_1$] (v1) {};
        \draw (2,0.5) node[vertex,label=right:$v_2$] (v2) {};
        \draw (2,-0.5) node[vertex,label=right:$v_3$] (v3) {};
        \draw (2,-1.5) node[vertex,label=below:$v_4$] (v4) {};
        \draw (4,1.5) node[vertex,label=right:$x_1$] (x1) {};
        \draw (4,0.5) node[vertex,label=right:$x_2$] (x2) {};
        \draw (4,-0.5) node[vertex,label=right:$x_3$] (x3) {};
        \draw (4,-1.5) node[vertex,label=right:$x_4$] (x4) {};

        \draw (u)--(v1) (u)--(v2) (u)--(v3) (u)--(v4)
	      (v1)--(v2) (v2)--(v3) 
              (v1)--(x1) (v3)--(x2) (v4)--(x3);
        \draw[very thick, double] (v4)--(x4);
       \end{tikzpicture}
       \caption{The case $p_3'=3$ 
          ($P_3'=\set{v_1,v_2,v_3}$)
	  may lead to just two good degree reductions.
         \label{TwoGoodRed}
       }
\end{figure}

If $p_3' = 3$, again either there is a gain of $2 h_4$ in a nonregular branch, 
or each super 2-reduction occurs in a regular branch. 
The 3 vertices in $P'_3$ have at least 2 neighbors in $N^2$,
or else these neighbors, along with $P_3 \setminus P'_3$, 
would form a cut of size 2 or smaller.
Each of these neighbors has its degree reduced, 
and thus must get deleted from a regular $F_i$, 
for a gain of at least $2 a_3$.
So there is either a gain of $2 h_4$ in a nonregular branch,
or a gain of $2a_3$ altogether in one or two regular branches.
(We cannot hope to claim $3 h_4$ or $3 a_3$, 
per the example in Figure~\ref{TwoGoodRed}.)

If $p'_3 = 4$, we claim that in the two branches together there are 
at least 4 good degree reductions on vertices in $N^2$ and $N^3(u)$.
Each contributes a gain of at least $h_4$ if it is in a nonregular branch,
$a_3$ in a regular branch.
Each vertex in $N^2$ undergoes a good degree reduction 
in one branch or the other, 
so if $|N^2| \geq 4$ we are done.
Since there can be no 2-cut, we may otherwise assume that $|N^2|=3$.
Since (in $F$) every vertex in $N(u)$ has degree~3,
there is an even number of edges between $N(u)$ and $N^2$,
thus there are at least 4 such edges.
Since each vertex in $N^2$ has an edge from $N(u)$, 
there must be two such edges incident on one vertex $x_1 \in N^2$,
and one edge each incident on the other vertices $x_2, x_3 \in N^2$.
Again we guaranteed 4 good degree reductions unless $x_1$ has degree 3
\emph{and} undergoes both of its reductions in one branch
(so that degree 3 to 2 is a good reduction, but 2 to 1 is not).
In that case, though, $x_1$ has degree~1, 
its remaining neighbor must be in $N^3(u)$
(otherwise $\set{x_1,x_2}$ is a 2-cut),
and 1-reducing on $x_1$ gives a good degree reduction on that neighbor. 
So there is a total gain of $4 h_4$ in a nonregular branch and
$4 a_3$ in a regular branch.

By convexity, the elementwise average of two pairs of splitting numbers
is a constraint dominated by one or the other,
so it suffices to write down the extreme constraints,
with all the gain from super 2-reductions
given to a single nonregular or regular branch.

Before counting the super 2-reduction gains, 
if $F_i$ is nonregular
the measure decrease $\mu(F)-\mu(F_i)$ is at least
\begin{align}
 \Delta_{\nonreg}(p_3, p_3'', p_4) 
  & := a_4 + \sum_{i=3}^4 p_i h_i + \ceil{\tfrac{p_3''}{2}} (w_h - w_e) , \label{deg4nonregBase}
\intertext{and if $F_i$ is $4$-regular, at least}
 \Delta_{\reg}(p_3, p_3'', p_4) 
  & := a_4 + \sum_{i=3}^4 p_i a_i + \ceil{\tfrac{p_3''}{2}} (w_h - w_e) - R_4 . \label{deg4regBase}
\end{align}
The super 2-reductions give an additional gain, in a nonregular branch, 
of at least
\begin{align}
 g_\nonreg & := \floor{\tfrac{p_3'+2}{3}} 2 h_4 , \label{deg4nonregBonus}
\intertext{and in a regular branch, at least}
 g_\reg & := \parens{ \floor{\tfrac{p_3'}2} + \floor{\tfrac{p_3'}3} + \floor{\tfrac{p_3'}4} } a_3, \label{deg4regBonus}
\end{align}
where the tricky floor and ceiling expressions 
are just a way of writing an explicit expression
convenient for passing to the nonlinear solver.
The constraints arising from splitting on a vertex of degree 4
with at least one neighbor of degree 3 are thus dominated by
the following, 
taken over $p_3' + p_3'' + p_4 = 4$, with $p_4 \leq 3$ and $p_3=p_3'+p_3''$:
\begin{align}
 (\Delta_\nonreg &, \Delta_\nonreg + g_\nonreg) \label{deg41}
,\\
 (\Delta_\nonreg &, \Delta_\reg + g_\reg) \label{deg42}
,\\
 (\Delta_\reg &, \Delta_\nonreg + g_\nonreg) \label{deg43}
,\\
 (\Delta_\reg &, \Delta_\reg + g_\reg) \label{deg44}
.
\end{align}

\section{Instances of degree 5}
\label{subsec:maxdeg5}

This section considers formulas of maximum degree~$5$.
As an overview, if there is a $3$-cut isolating 
a set $S$ with $6$ or more vertices 
the algorithm splits on any vertex in the cut.
Otherwise, the algorithm chooses a vertex $u$ of degree $5$ with 
--- if possible --- at least
one neighbor of degree at most~$4$,
and splits on~$u$
either as was done in the degree-4 case,
or using clause-learning splitting (see Lemma~\ref{clause}).
We use clause learning when the neighbors of $u$ have high degrees,
because clause learning sets many variables in $N(u)$,
and this is most effective when the degrees are large (since $a_i \ge a_{i-1}$).
We use normal splitting when the neighbors have low degrees,
because setting $u$ reduces their degrees,
and this is effective when the degrees are small
($h_i \le h_{i+1}$, 
with an additional bonus in super 2-reductions for a degree-3 variable).
(This is also why we always prefer to split on 
vertices of maximum degree with neighbors of low degree, 
and why the regular cases need special attention.)

\myitem{3-cut} \label{cut53}
There is a $3$-cut $C = \{ x_1,x_2,x_3 \}$ isolating a set $S$ 
of vertices such that
$6 \leq |S| \leq 10$ and $S$ contains at least one vertex of degree~$5$.
Splitting on the cut vertex $x_1$ leaves 
constraint graphs where $\{x_2,x_3\}$ form a $2$-cut. 
Thus $S \cup \{x_1\}$ are removed from both resulting instances ($a_5+6a_3$),
a neighbor of $x_1$ outside $S \cup C$ has its degree reduced ($h_5$), 
a heavy edge $x_2 x_3$ appears (in the worst case) but at least $2$ half-edges
incident on $x_2$ and $x_3$ disappear ($-w_h+w_e$). 
Additionally, the resulting instances may become $5$-regular ($-R_5$). 
So, the splitting number is at most
\begin{align}
 \big( a_5 + 6 a_3 + h_5 -w_h + w_e - R_5, 
       \, a_5 + 6 a_3 + h_5 -w_h + w_e - R_5 \big) .
 \label{splittingCut53}
\end{align}
\caseend

In light of reduction~\ref{cut53} 
we may henceforth assume that each degree-5 variable $u$ has $|N^2(u)| \ge 4$.

\myitem{5-regular} \label{CaseReg5}
If every vertex has degree~$5$, 
the same analysis as for 4-regular instances (reduction~\ref{CaseReg4}, 
constraints \eqref{splittingReg41} and~\eqref{splittingReg42})
gives a splitting number which is at most one of the following:
\begin{align}
 ( 6 a_5 &, 6 a_5 ) , \label{splittingReg51}
 \\
 (a_5 + 5 h_5 + R_5 &, a_5 + 5 h_5 + R_5 ) . \label{splittingReg52}
\end{align}
\caseend

Otherwise, let $u$ be a degree-5 vertex with a minimum number
of degree-5 neighbors,
and as usual let $p_i$ be the number of degree-$i$ neighbors of $u$
(since the instance is not regular, $p_5 < 5$).
Let $H:=\chi(u \text{ is incident to a heavy edge})$. 
Depending on the values of $H$ and $p_i$ we will use 
either 
regular 2-way splitting (reduction~\ref{CaseNonReg52}) or 
clause-learning 3-way splitting (reduction~\ref{CaseNonReg51}).

\newcommand{\NN}{N^2}

\myitem{5-nonregular, 2-way splitting} \label{CaseNonReg52}
$H=1$ or $p_3 \ge 1$ or $p_5 \le 2$.

In this 
case 
we use the usual $2$-way splitting, setting $u$ to $0$ and to~$1$,
and simplifying to obtain $F_0$ and $F_1$.
If $F_i$ is not regular, the measure decrease $\mu(F)-\mu(F_i)$
is at least
$a_5 + \sum_{i=3}^5 p_i h_i + H (w_h-w_e)$,
and if $F_i$ is $5$-regular,
it is at least $a_5 + \sum_{i=3}^5 p_i a_i + H (w_h-w_e) - R_5$.
Thus if both branches are regular the splitting number is at most
\begin{align}
 \big( a_5 + \ssum_{i=3}^5 p_i a_i + H (w_h-w_e) - R_5
    &, \;
       a_5 + \ssum_{i=3}^5 p_i a_i + H (w_h-w_e) - R_5  \big) ,
 \label{splittingNonreg51} 
 \intertext{and if one branch is regular and one nonregular, at most}
 \big( a_5 + \ssum_{i=3}^5 p_i a_i + H (w_h-w_e) - R_5
    &, \;
       a_5 + \ssum_{i=3}^5 p_i h_i + H (w_h-w_e)        \big) .
    \label{splittingNonreg52} 
\end{align}
If both branches are nonregular, we use that if $p_3 \ge 1$, 
any degree-3 neighbor of $u$ either has a heavy edge not incident to~$u$,
giving an additional measure reduction of at least $w_h-w_e$,
or in at least one branch may be super 2-reduced,
for a measure reduction of at least $2 h_5$. 
(The latter requires a justification we give explicitly, 
although Lemma~\ref{1lemma} could be invoked.
At the start of the first super 2-reduction, 
every vertex has degree 2 or more.
Each of the two ``legs'' of the super 2-reduction 
propagates through a [possibly empty] chain of degree-2 vertices
before terminating either in a good degree reduction 
or by meeting a vertex that was reduced to degree 1 by the other leg.
In the latter case all the vertices involved had degree~2,
thus were neighbors of $u$ originally of degree~3;
also, there must have been at least three of them to form a cycle, 
and the remaining 2 or fewer vertices in $N(u)$ 
contradict the assumption that $F$ was simplified.)
Thus, the splitting number is at most
\begin{align}
 \big( a_5 + \ssum_{i=3}^5 p_i h_i + H (w_h-w_e) + \chi(p_3 \ge 1) 2 h_5 
       &, \;
       a_5 + \ssum_{i=3}^5 p_i h_i + H (w_h-w_e)         \big) \text{ or}
  \label{splittingNonreg53} 
\end{align}
\begin{align}
 \big( 
   & a_5 + \ssum_{i=3}^5 p_i h_i + H (w_h-w_e) + \chi(p_3 \ge 1) (w_h-w_e) 
       , \notag \\
   & a_5 + \ssum_{i=3}^5 p_i h_i + H (w_h-w_e) + \chi(p_3 \ge 1) (w_h-w_e)  
 \big) .
  \label{splittingNonreg54} 
\end{align}
\caseend

\myitem{5-nonregular, clause learning} \label{CaseNonReg51}
$H=0$ and $p_3=0$ and $p_5 \in \set{3,4}$.

Let $v$ be a degree~$5$ (degree $5$ in $G$) neighbor of $u$ 
with a minimum number of degree-$5$ neighbors in $\NN := N^2(u)$.
The clause learning splitting (see Lemma~\ref{clause})
will set $u$ in the first branch,
$u$ and $v$ in the second branch,
and all of $N[u]$ in the third branch. 
In each of the $3$ branches, 
the resulting instance could become $5$-regular or not.

In the \emph{first branch}, the measure of the instance decreases by at least
\begin{align}
 \Delta_{51} := \min
 \begin{cases}
  a_5 + \sum_{i=4}^5 p_i h_i &\text{($5$-nonregular case), or}\\
  a_5 + \sum_{i=4}^5 p_i a_i - R_5 &\text{($5$-regular case).}
 \end{cases}
 \label{f1}
\end{align}

In the analysis of the second and third branches 
we distinguish between the case where $v$ 
has at most one neighbor of degree $5$ in~$\NN$,
and the case where $v$ (and thus every degree-$5$ neighbor of~$u$)
has at least two neighbors of degree $5$ in $\NN$.

In the \emph{second branch}, 
if $v$ has at most one neighbor of degree $5$ in~$\NN$,
the measure of the instance decreases by at least
\begin{align}
 \Delta_{52}^1 := \min
 \begin{cases}
  a_5 + \sum_{i=4}^5 p_i h_i + a_4 + 3 h_4 + h_5 
     &\text{($5$-nonregular case), or}\\
  a_5 + \sum_{i=4}^5 p_i a_i -R_5 
     &\text{($5$-regular case).}
 \end{cases}
 \label{f2a}
\end{align}
(The degree reductions $3 h_4 + h_5$ from the nonregular case 
do not appear in the regular case because they 
may pertain to the same vertices
as the deletions $\sum p_i a_i$.)

If $v$ has at least two neighbors of degree $5$ in $\NN$, 
the measure decreases by at least
\begin{align}
 \Delta_{52}^2 := \min
 \begin{cases}
  a_5 + \sum_{i=4}^5 p_i h_i + a_4 + 4 h_5 &\text{($5$-nonregular case), or}\\
  a_5 + \sum_{i=4}^5 p_i a_i + 2 a_5 -R_5 &\text{($5$-regular case).}
 \end{cases}
 \label{f2b}
\end{align}

In the \emph{third branch}, 
first take the case where $v$ has at most one neighbor of degree $5$ in $\NN$.
Since $|\NN| \geq 4$, there are at least 4 good degree reductions
on vertices in $\NN$.
If the instance becomes regular, 
this implies a measure decrease of at least $4 a_3$.
If the instance remains nonregular,
this is a measure reduction of at least $4 h_5$,
and we now show that if $p_5=4$ then
there is a fifth good degree reduction.
We argue this just as the 4-nonregular case 
(Section~\ref{CaseNonReg4}) with $p_3'=4$;
we could alternatively apply Lemma~\ref{1lemma}.
If $|\NN|=5$ the desired $5 h_5$ is immediate.
Otherwise, $|\NN|=4$,
the number of edges between $N(u)$ and $\NN$ is at least~4, 
and odd (from $p_5=4$ and $p_4=1$, recalling that $p_3=0$),
so $|N(u) \times \NN| \geq 5$.
At least one edge incident on each vertex in $\NN$ gives a good degree
reduction, and we fail to get a fifth such reduction only if 
the fifth edge is incident on a vertex $x \in \NN$ of degree~3,
leaving it with degree~1.  
But in that case the remaining neighbor of $x$ must be in $N^3(u)$
(otherwise $\NN \setminus x$ is a 3-cut,
a contradiction by reduction~\ref{cut53}),
and 1-reducing $x$ gives the fifth good degree reduction.
Thus the measure decreases by at least
\begin{align}
 \Delta_{53}^1 := \min
 \begin{cases}
  a_5 + \sum_{i=4}^5 p_i a_i + 4 h_5 + \chi(p_5=4) h_5 
       &\text{($5$-nonregular case), or}\\
  a_5 + \sum_{i=4}^5 p_i a_i + 4 a_3 -R_5 
       &\text{($5$-regular case).}
 \end{cases}
 \label{f3a}
\end{align}

Otherwise, in the third branch, 
$v$ has at least two neighbors of degree $5$ in~$\NN$.
For the regular case we simply note that each vertex in $N^2$
has its degree reduced and must be deleted,
$N^2$ has at least four vertices of which at least two are of degree~5,
for a measure reduction of at least $2a_5+2a_3$.
We now address the nonregular case.
Letting $P_5$ be the set of degree-5 vertices in $N(u)$
(so $|P_5|=p_5$),
by definition of $v$ every vertex in $P_5$ has at least 
two degree-5 neighbors in~$\NN$.
Let $R \subseteq \NN$ be the set of degree-5 vertices in $\NN$
adjacent to~$P_5$,
and let $E_5= E \cap (P_5 \times R)$ be the set of edges between $P_5$ and $R$.
There is one last case distinction, according to the value of~$p_5$.
If $p_5=3$ there are at least 6 good degree reductions:
$|E_5|=6$, 
each vertex in $R$ has at most $|P_5|=3$ incident edges from $E_5$,
and thus each such incidence results in a good degree reduction
(the vertex degree is reduced at most from 5 to 4 to 3 to 2).
Here we have $6 h_5$.

If $p_5=4$ we claim that the good degree reductions
amount to at least $\min \set{8 h_5, 5h_5 + h_4 + h_3}$.
By default the 8 edges in $E_5$ all generate good degree reductions,
with fewer only if some of the degree-5 vertices in $R$ have 
more than 3 incident edges from $E_5$.  
The ``degree spectrum'' on $R$ is thus a partition of 8
(the number of incident edges)
into $|R|$ parts, where no part can be larger than $|P_4|=4$.
If the partition is $4+4$ this means two reductions that are not good
($2h_2$), but then this implies that $|R|=2$,
and the other two vertices in $N^2 \setminus R$
also have their degrees reduced, 
restoring the total of 8 good reductions.
If the partition has exactly one $4$, on a vertex $r \in R$,
then just one of the 8 degree reductions is not good, 
and the 7 good reductions include those on $r$,
thus giving a measure reduction of at least $5h_5+h_4+h_3$.

Considering the difference, which we will denote $g_{p_5=4}$, 
between these guaranteed measure decreases 
and the guarantee of $6 h_5$ when $p_5=3$,
we constrain
\begin{align}
 g_{p_5=4} & \le 8 h_5 - 6 h_5 = 2 h_5, \label{g5a}\\
 g_{p_5=4} & \le (5 h_5 +h_4+h_3)-6h_5 = -h_5+h_4 + h_3. \label{g5b}
\end{align}
and we obtain a measure reduction of at least
\begin{align}
 \Delta_{53}^2 := \min
 \begin{cases}
  a_5 + \sum_{i=4}^5 p_i a_i + 6 h_5 + \chi(p_4=1) g_{p_5=4} &\text{($5$-nonregular case), or}\\
  a_5 + \sum_{i=4}^5 p_i a_i + 2 a_5 + 2 a_3 -R_5 &\text{($5$-regular case).}
 \end{cases}
 \label{f3b}
\end{align}

Wrapping up this reduction, 
the case that $v$ has at most 1 degree-5 neighbor in $N$,
or at least two such neighbors,
respectively impose the constraints (splitting numbers)
\begin{align}
 ( \Delta_{51}, & \Delta_{52}^1, \Delta_{53}^1 ) \text{ and} \label{clauseLearning51}
\\
 ( \Delta_{51}, & \Delta_{52}^2, \Delta_{53}^2 ) . \label{clauseLearning52}
\end{align}
\caseend

\section{Instances of degree 6}
\label{subsec:maxdeg6}

This section considers formulas of maximum degree $6$.
The algorithm chooses a vertex $u$ of degree $6$ with 
--- if possible --- at least
one neighbor of lower degree, and splits on $u$ 
by setting it to $0$ and~$1$.

\myitem{6-regular} \label{CaseReg6}
If every vertex has degree~$6$, 
the same analysis as for regular instances of degree $4$ 
gives a splitting number which is at least one of the following:
\begin{align}
 ( 7 a_6 &, 7 a_6 ) , \label{splittingReg61}
 \\
 ( a_6 + 6 h_6 + R_6 &, a_6 + 6 h_6 + R_6 ) . \label{splittingReg62}
\end{align}
\caseend

\myitem{6-nonregular} \label{CaseNonReg61}
Vertex $u$ has at least one neighbor of degree at most $5$.

It is straightforward that the splitting number is at least as large
as one of the following
(only distinguishing if the instance becomes $6$-regular or not):
\begin{align}
 \bigg( a_6 + \sum_{i=3}^6 p_i h_i  &, \; 
        a_6 + \sum_{i=3}^6 p_i h_i \bigg) , \label{splittingNonreg61}
\\
 \bigg( a_6 + \sum_{i=3}^6 p_i a_i - R_6 &,  \;
        a_6 + \sum_{i=3}^6 p_i a_i - R_6 \bigg) . \label{splittingNonreg62}
\end{align}
\caseend

\section{Tuning the bounds} \label{tuning}
For any values of $w_e$ and $w_h$ satisfying the constraints 
we have set down,
we have shown that any
\mc instance $F$ is solved in time 
$\Ostar{\twopow{|E| w_e + |H| w_h}}$.

For a given instance $F$, the
running-time bound is best for the feasible values of 
$w_e$ and $w_h$ which minimize $|E| w_e + |H| w_h$.
As usual taking $|E|=(1-p)m$ and $|H|=pm$,
this is equivalent to minimizing 
\begin{align}
  (1-p) w_e + p w_h , \label{objective}
\end{align}
allowing us to obtain a 1-parameter family of running-time bounds
--- pairs $(w_e,w_h)$ as a function of~$p$ ---
tuned to a formula's fraction of conjunctive and general 2-clauses.

Reiterating, if a formula's ``p'' value is $p(F) = |H|/(|E|+|H|)$,
and if minimizing \eqref{objective} 
for a given $p$ gives a pair $(w_e,w_h)(p)$,
then the optimal bound for formula $F$ is the one given by $(w_e,w_h)(p(F))$,
but for \emph{any} $(w_e,w_h)(p)$,
the running-time bound $\Ostar{\twopow{|E|w_e+|H|w_h}}$
is valid for every formula $F$, even if $p \neq p(F)$.
This is simply because every such pair $(w_e,w_h)$ is a feasible
solution of the nonlinear program,
even if it is not the optimal solution for the 
appropriate objective function.

For cubic instances, minimizing \eqref{objective} with $p$ small 
gives 
$w_e \approx 0.10209$ and $w_h \approx 0.23127$,
while minimizing with $p$ close to 1 gives
$w_e = w_h = 1/6$ 
(the tight constraints are all linear, so the solution is rational),
matching the best known polynomial space running time for
general instances of \mc (see~\cite{faster}).
It appears that the first result is obtained for all $p \leq 1/2$
and the second for all $p>1/2$.

For instances of degrees 4, 5, and 6 or more,
the results of minimizing with various values of $p$
are shown in Table~\ref{tab:runtimes},
and the most interesting of these is surely 
that of degree 6 or more (the general case).
Here, taking $p$ small gives 
$w_e \approx 0.15820$ and $w_h \approx 0.31174$. 
For instances of \mts this gives a running-time bound 
of $\Ostar{\twopow{0.1582 m}}$ or $\Ostar{\twopow{m/6.321}}$,
improving on the best bound previously known,
giving the same bound for mixtures of OR and AND clauses,
and giving nearly as good run times when 
a small fraction of arbitrary integer-weighted clauses are mixed in.
We observe that any $p \geq 0.29$ leads to
$w_e = w_h = 0.19$ 
(as for cubic case with $p>1/2$, 
the tight constraints are linear, so the value is rational),
matching the best known bound (for polynomial-space algorithms) of
$\Ostar{\twopow{0.19 m}}$ from~\cite{faster}.
Figure~\ref{p6plot} shows the values of $w_e$, $w_h$,
and the objective $(1-p)w_e + (p)w_h$,
as a function of~$p$.
Numerically, the values $w_e$ and $w_h$ meet for some value of $p$
between $0.2899$ and $0.29$.

\section*{Acknowledgment}
The authors are very grateful to Alex Scott for initiating
this project and contributing some of the first key ideas.

\providecommand{\bysame}{\leavevmode\hbox to3em{\hrulefill}\thinspace}
\providecommand{\MR}{\relax\ifhmode\unskip\space\fi MR }
\providecommand{\MRhref}[2]{%
  \href{http://www.ams.org/mathscinet-getitem?mr=#1}{#2}
}
\providecommand{\href}[2]{#2}

\newpage
\appendix
\section*{Appendix: Convex Program for Computing the Optimal Weights}

Below we show, in AMPL notation, the objective function and
all the constraints of the mathematical program we solve
to optimize an algorithm for hybrid instances with
a fraction \texttt{p} of non-simple clauses.
Constraints are annotated the numbers of the corresponding inequalities
in the paper's body.
The parameter \texttt{margin} is the ``$\epsilon$'' discussed in
Section~\ref{solver} to ensure that a solution is truly feasible
even in the face of finite-precision arithmetic.

\begin{Verbatim}[commandchars=\\\|\%]
# Max 2-Sat and Max 2-CSP

# maximum degree
param maxd integer >=3;
# fraction of non-simple clauses
param p;
param margin;
set DEGREES := 0..maxd;
# weight for edges
var We >= 0;
# weight for degree reductions from degree at most i
var h {DEGREES} >= 0;
# vertex of degree i + i/2 surrounding half edges
var a {DEGREES};
# weight for heavy edges
var Wh;
# Regular weights
var R4 >= 0; \hfill\eqref|Rpos%
var R5 >= 0; \hfill\eqref|Rpos%
var R6 >= 0; \hfill\eqref|Rpos%
# additional degree reductions in the 3rd branch (nonregular)
# of the clause learning branching for p5=4 vs p5=3
var nonreg53;
# change in measure for the 3 branches
# 1st argument is the nb of deg-4 nbs of u
# 2nd argument distinguishes (if present) if v has at most 1 deg-5 nb in N^2 (1)
#     or at least 2 (2)
set TWO := 1..2;
var f1 {TWO};
var f2 {TWO,TWO};
var f3 {TWO,TWO};
var D4r {0..4, 0..4};
var D4n {0..4, 0..4};
var g4r {0..4};
var g4n {0..4};

# analysis in terms of the number of edges
minimize Obj: (1-p)*We + p*Wh;

# Some things we know
subject to Known:
  a[0] = 0; \hfill\eqref|a00%

# Constrain W values non-positive
subject to Wnonpos {d in DEGREES : d>=1}:
  a[d] - d*We/2 <= 0 - margin; \hfill\eqref|ai%\eqref|wneg%

# a[] value positive
subject to MeasurePos {d in DEGREES : d>=1}:
  a[d] >= 0 + margin; \hfill\eqref|apos%

# Intuition: weight for heavy edges >= weight for light edges
subject to HeavyEdge:
  We - Wh <= 0 - margin; \hfill\eqref|whwe%

# collapse parallel edges 
subject to parallel {d in DEGREES : d >= 3}:
   Wh - We - 2*a[d] + 2*a[d-1] <= 0 - margin; \hfill\eqref|heavy%

# decomposable edges
subject to Decomposable {d in DEGREES : d >= 1}:
   - a[d] + a[d-1] <= 0 - margin; \hfill\eqref|decomposable%

# constraints for the values of h[]
subject to hNotation {d in DEGREES, i in DEGREES : 3 <= i <= d}:
  h[d] - a[i] + a[i-1] <= 0 - margin; \hfill\eqref|h3def%\eqref|halfred%
 
#######################################
# constraints for cubic
#######################################
 
# 3-cut
subject to Cut3:
  2*2^(-5*a[3] - 2*h[3]) <= 1 - margin; \hfill\eqref|3cut%

# Independent neighborhood
subject to Indep {q in 0..3}:
  2^(-a[3] - 3*h[3] -q*(Wh-We)) + 2^(-a[3] -3*h[3] - q*(Wh-We) - 2*(3-q)*h[3])
  <= 1 - margin; \hfill\eqref|indep3%

# One edge in neighborhood
subject to OneEdge1:
  2^(-5*a[3]-h[3]) + 2^(-5*a[3] -3*h[3]) <= 1 - margin; \hfill\eqref|edge32%

subject to OneEdge2:
  2^(-5*a[3] - h[3] - Wh + We) + 2^(-5*a[3] - h[3] - Wh + We) <= 1 - margin; \hfill\eqref|edge31%

#######################################
# constraints for degree 4
#######################################

# 4-regular

# regular becomes nonregular
subject to Regular41:
   2* 2^(-a[4] - 4*h[4]-R4) <= 1 - margin; \hfill\eqref|splittingReg42%

# regular becomes regular
subject to Regular42:
   2* 2^(-5*a[4]) <= 1 - margin; \hfill\eqref|splittingReg41%

# 4 non-regular

subject to 4nonregularBase 
  {p3p in 0..4, p3pp in 0..4, p4 in 0..3: p3p+p3pp+p4=4}: 
  D4n[p3p,p3pp] = -a[4] -(p3p+p3pp)*h[3] -p4*h[4] -ceil(p3pp/2)*(Wh-We); \hfill\eqref|deg4nonregBase%

subject to 4regularBase 
  {p3p in 0..4, p3pp in 0..4, p4 in 0..3: p3p+p3pp+p4=4}: 
  D4r[p3p,p3pp] = -a[4] -(p3p+p3pp)*a[3] -p4*a[4] -ceil(p3pp/2)*(Wh-We) +R4; \hfill\eqref|deg4regBase%

subject to 4nonregularBonus 
  {p3p in 0..4, p3pp in 0..4, p4 in 0..3: p3p+p3pp+p4=4}: 
  g4n[p3p] = - floor((p3p+2)/3) * (2*h[4]); \hfill\eqref|deg4nonregBonus%

subject to 4regularBonus 
  {p3p in 0..4, p3pp in 0..4, p4 in 0..3: p3p+p3pp+p4=4}: 
  g4r[p3p] = - (floor(p3p/2)+floor(p3p/3)+floor(p3p/4)) * a[3]; \hfill\eqref|deg4regBonus%

subject to Nonregular41 {p3p in 0..4, p3pp in 0..4, p4 in 0..3: p3p+p3pp+p4=4}:
  2^(D4n[p3p,p3pp]) + 2^(D4n[p3p,p3pp] + g4n[p3p]) 
  <= 1 - margin; \hfill\eqref|deg41%

subject to Nonregular42 {p3p in 0..4, p3pp in 0..4, p4 in 0..3: p3p+p3pp+p4=4}:
  2^(D4n[p3p,p3pp]) + 2^(D4r[p3p,p3pp] + g4r[p3p]) 
  <= 1 - margin; \hfill\eqref|deg42%

subject to Nonregular43 {p3p in 0..4, p3pp in 0..4, p4 in 0..3: p3p+p3pp+p4=4}:
  2^(D4r[p3p,p3pp]) + 2^(D4n[p3p,p3pp] + g4n[p3p]) 
  <= 1 - margin; \hfill\eqref|deg43%

subject to Nonregular44 {p3p in 0..4, p3pp in 0..4, p4 in 0..3: p3p+p3pp+p4=4}:
  2^(D4r[p3p,p3pp]) + 2^(D4r[p3p,p3pp] + g4r[p3p]) 
  <= 1 - margin; \hfill\eqref|deg44%

#######################################
# constraints for degree 5
#######################################

# 3-cut for degree 5
subject to Cut5_3:
   2* 2^(-a[5] - 6*a[3] + R5 +(Wh-We)) <= 1 - margin; \hfill\eqref|splittingCut53%

# 5-regular

#  regular becomes nonregular
subject to Regular51:
   2* 2^(-a[5] - 5*h[5]-R5) <= 1 - margin; \hfill\eqref|splittingReg51%

#  regular stays regular
subject to Regular52:
   2* 2^(-6*a[5]) <= 1 - margin; \hfill\eqref|splittingReg52%

# 5 non-regular

# clause learning

# first branch
subject to Cf1 {p4 in 1..2, p5 in 3..4: p4+p5=5}:
   f1[p4] >= -a[5]-p4*h[4]-p5*h[5]; \hfill\eqref|f1%
subject to Cf1reg {p4 in 1..2, p5 in 3..4: p4+p5=5}:
   f1[p4] >= -a[5]-p4*a[4]-p5*a[5]+R5; \hfill\eqref|f1%

# second branch, v has at most 1 deg-5 neighbor in N^2
subject to Cf2a {p4 in 1..2, p5 in 3..4: p4+p5=5}:
   f2[p4,1] >= -a[5]-p4*h[4]-p5*h[5]-a[4]-3*h[4]-h[5]; \hfill\eqref|f2a%
subject to Cf2areg {p4 in 1..2, p5 in 3..4: p4+p5=5}:
   f2[p4,1] >= -a[5]-p4*a[4]-p5*a[5]+R5; \hfill\eqref|f2a%

# second branch, v (and all other deg-5 nbs of u) has at least 2 deg-5 nbs in N^2
subject to Cf2b {p4 in 1..2, p5 in 3..4: p4+p5=5}:
   f2[p4,2] >= -a[5]-p4*h[4]-p5*h[5]-a[4]-4*h[5]; \hfill\eqref|f2b%
subject to Cf2breg {p4 in 1..2, p5 in 3..4: p4+p5=5}:
   f2[p4,2] >= -a[5]-p4*a[4]-p5*a[5]-2*a[3]+R5; \hfill\eqref|f2b%

# additional degree reductions in the 3rd branch (nonregular) for p5=4 vs p5=3
subject to addDegRedNR53_1:
  nonreg53 <= 2*h[5]; \hfill\eqref|g5a%
subject to addDegRedNR53_2:
  nonreg53 <= h[4]+h[3]-h[5]; \hfill\eqref|g5b%

# third branch, v has at most 1 deg-5 neighbor in N^2
subject to Cf3a {p4 in 1..2, p5 in 3..4: p4+p5=5}:
   f3[p4,1] >= -a[5]-p4*a[4]-p5*a[5]-(4+((4*p4+5*p5-5) mod 2))*h[5]; \hfill\eqref|f3a%
subject to Cf3areg {p4 in 1..2, p5 in 3..4: p4+p5=5}:
   f3[p4,1] >= -a[5]-p4*a[4]-p5*a[5]-4*a[3]+R5; \hfill\eqref|f3a%

# third branch, v (and all other deg-5 nbs of u) has at least 2 deg-5 nbs in N^2
subject to Cf3b {p4 in 1..2, p5 in 3..4: p4+p5=5}:
   f3[p4,2] >= -a[5]-p4*a[4]-p5*a[5]-6*h[5]-floor(p5/4)*nonreg53; \hfill\eqref|f3b%
subject to Cf3breg {p4 in 1..2, p5 in 3..4: p4+p5=5}:
   f3[p4,2] >= -a[5]-p4*a[4]-p5*a[5]-2*a[3]-2*a[5]+R5; \hfill\eqref|f3b%

# the clause learning splitting
subject to Nonregular5cl {p4 in 1..2, nb5 in 1..2}:
   2^(f1[p4]) + 2^(f2[p4,nb5]) + 2^(f3[p4,nb5]) <= 1; \hfill\eqref|clauseLearning51%

# 2-way splitting

# 2-way splitting, non-reg in both branches, if p3>0, then additional heavy edge
subject to Nonregular51a {p3 in 0..5, p4 in 0..5, 
                          p5 in 0..4, H in 0..1: p3+p4+p5=5
                          and ((H=1) or (p5 < 3 or p3>0))}:
   2* 2^(-a[5] - p3*h[3] - p4*h[4] - p5*h[5] -H*(Wh-We) -ceil(p3/5)*(Wh-We))
   <= 1 - margin; \hfill\eqref|splittingNonreg54%

# 2-way splitting, non-reg in both branches, if p3>0, then additional super-2
subject to Nonregular51b {p3 in 0..5, p4 in 0..5, p5 in 0..4, 
                          H in 0..1: p3+p4+p5=5
                          and ((H=1) or (p5 < 3 or p3>0))}:
   2^(-a[5] - p3*h[3] - p4*h[4] - p5*h[5] -H*(Wh-We) -ceil(p3/5)*2*h[5])
 + 2^(-a[5] - p3*h[3] - p4*h[4] - p5*h[5] -H*(Wh-We))
 <= 1 - margin; \hfill\eqref|splittingNonreg53%

# 2-way splitting, becomes reg in both branches
subject to Nonregular52 {p3 in 0..5, p4 in 0..5, p5 in 0..4, 
                         H in 0..1: p3+p4+p5=5
                         and ((H=1) or (p5 < 3 or p3>0))}:
   2* 2^(-a[5] - p3*a[3] - p4*a[4] - p5*a[5] -H*(Wh-We) + R5) <= 1 - margin; \hfill\eqref|splittingNonreg51%

# 2-way splitting, becomes reg in 1 branch
subject to Nonregular52b {p3 in 0..5, p4 in 0..5, p5 in 0..4, 
                          H in 0..1: p3+p4+p5=5
                          and ((H=1) or (p5 < 3 or p3>0))}:
   2^(-a[5] - p3*a[3] - p4*a[4] - p5*a[5] -H*(Wh-We) + R5)
 + 2^(-a[5] - p3*h[3] - p4*h[4] - p5*h[5] -H*(Wh-We))
 <= 1 - margin; \hfill\eqref|splittingNonreg51%

#######################################
# constraints for degree 6
#######################################

# 6-regular

# regular becomes nonregular
subject to Regular61:
   2* 2^(-a[6] - 6*h[6]-R6) <= 1 - margin; \hfill\eqref|splittingReg62%

# regular stays regular
subject to Regular62:
   2* 2^(-7*a[6]) <= 1 - margin; \hfill\eqref|splittingReg61%

# 6 non-regular

# nonregular stays nonregular
subject to Nonregular61 {p3 in 0..6, p4 in 0..6, p5 in 0..6, p6 in 0..5: 
                          p3+p4+p5+p6=6}:
  2* 2^(-a[6] - p6*h[6] - p5*h[5] - p4*h[4] - p3*h[3]) <= 1 - margin; \hfill\eqref|splittingNonreg61%

# nonregular becomes regular
subject to Nonregular62 {p3 in 0..6, p4 in 0..6, p5 in 0..6, p6 in 0..5: 
                         p3+p4+p5+p6=6}:
  2* 2^(-a[6] - p6*a[6] - p5*a[5] - p4*a[4] - p3*a[3] +R6) <= 1 - margin; \hfill\eqref|splittingNonreg62%
\end{Verbatim}

\end{document}